\documentclass[amssymb,nofootinbib,superscriptaddress,twocolumn,accepted=2018-04-18]{quantumarticle}
\usepackage[utf8]{inputenc}
\usepackage[english]{babel}
\usepackage[T1]{fontenc}
\usepackage{amsmath}
\usepackage{dsfont} 
\usepackage{graphicx} 
\usepackage{epsfig}
\usepackage{epstopdf}
\usepackage{color}

\usepackage[colorlinks]{hyperref}
\makeatletter
\newcommand\org@hypertarget{}
\let\org@hypertarget\hypertarget
\renewcommand\hypertarget[2]{%
  \Hy@raisedlink{\org@hypertarget{#1}{}}#2%
  }
\makeatother
\usepackage[figure,table]{hypcap}
\usepackage{enumerate}
\usepackage{float}
\hypersetup{
	bookmarksnumbered,
	pdfstartview={FitH},
	pdfpagemode={UseOutlines}}
\definecolor{darkgreen}{RGB}{50,190,50}
\definecolor{darkblue}{RGB}{0,0,190}
\definecolor{darkred}{RGB}{238,0,0}
\usepackage{soul}
\newcommand{\pr}{^{\prime}}
\newcommand{\prpr}{^{\prime\prime}}
\newcommand{\ket}[1]{\ensuremath{\left|\right.\!{#1}\!\left.\right\rangle}}

\newcommand{\bra}[1]{\ensuremath{\left\langle\right.\!{#1}\!\left.\right|}}

\newcommand{\scpr}[2]{\ensuremath{\left\langle\right.\hspace*{-1pt} #1 \hspace*{-1pt}\left|\right.\hspace*{-1pt} #2 \hspace*{-1pt}\left.\right\rangle}}

\newcommand{\nl}{\ensuremath{\hspace*{-0.5pt}}}
\newcommand{\nr}{\ensuremath{\hspace*{0.5pt}}}

\newcommand{\subtiny}[3]{\ensuremath{_{\hspace{#1 pt}\protect\raisebox{#2 pt}{\tiny{$ #3$}}}}}
\newcommand{\suptiny}[3]{\ensuremath{^{\hspace{#1 pt}\protect\raisebox{#2 pt}{\tiny{$ #3$}}}}}
\newcommand{\dv}{\ensuremath{|\hspace*{-1.3pt}|}}

\newcommand{\rplus}{\ensuremath{r_{\hspace{-0.9pt}\protect\raisebox{-0.6pt}{\tiny{$+$}}}}}
\newcommand{\rminus}{\ensuremath{r_{\hspace{-0.9pt}\protect\raisebox{-0.6pt}{\tiny{$-$}}}}}
\newcommand{\rpm}{\ensuremath{r_{\hspace{-0.9pt}\protect\raisebox{-0.6pt}{\tiny{$\pm$}}}}}

\newcommand{\rtplus}{\ensuremath{\tilde{r}_{\hspace{-0.9pt}\protect\raisebox{-0.6pt}{\tiny{$+$}}}}}
\newcommand{\rtminus}{\ensuremath{\tilde{r}_{\hspace{-0.9pt}\protect\raisebox{-0.6pt}{\tiny{$-$}}}}}
\newcommand{\rtpm}{\ensuremath{\tilde{r}_{\hspace{-0.9pt}\protect\raisebox{-0.6pt}{\tiny{$\pm$}}}}}

\newcommand{\expval}[1]{\ensuremath{\left\langle\right.\hspace*{-1pt} #1 \hspace*{-1pt}\left.\right\rangle}}
\newcommand{\comm}[2]{\ensuremath{\left[\right.\! #1 \,, #2 \!\left.\right]}}

\newcommand{\tr}{\textnormal{Tr}}
\newcommand{\djj}{d\kern-0.4em\char"16\kern-0.1em}

\renewcommand{\thesubsection}{\thesection.\Alph{subsection}}
\renewcommand{\thesubsubsection}{\Roman{section}.\Alph{subsection}.\arabic{subsubsection}}
\makeatletter
\renewcommand{\p@subsection}{}
\renewcommand{\p@subsubsection}{}
\makeatother

\begin{document}

\title{Precision and Work Fluctuations in Gaussian\\ Battery Charging}
\author{Nicolai Friis}
\email{nicolai.friis@univie.ac.at}
\affiliation{Institute for Quantum Optics and Quantum Information, Austrian Academy of Sciences, Boltzmanngasse 3, 1090 Vienna, Austria}
\affiliation{Institute for Theoretical Physics, University of Innsbruck, Technikerstra{\ss}e 21a, 6020 Innsbruck, Austria}
\orcid{0000-0003-1950-8640}
\author{Marcus Huber}
\email{marcus.huber@univie.ac.at}
\affiliation{Institute for Quantum Optics and Quantum Information, Austrian Academy of Sciences, Boltzmanngasse 3, 1090 Vienna, Austria}
\orcid{0000-0003-1985-4623}

\begin{abstract}
One of the most fundamental tasks in quantum thermodynamics is extracting energy from one system and subsequently storing this energy in an appropriate battery. Both of these steps, work extraction and charging, can be viewed as cyclic Hamiltonian processes acting on individual quantum systems. Interestingly, so-called passive states exist, whose energy cannot be lowered by unitary operations, but it is safe to assume that the energy of any not fully charged battery may be increased unitarily. However, unitaries raising the average energy by the same amount may differ in qualities such as their precision, fluctuations, and charging power. Moreover, some unitaries may be extremely difficult to realize in practice. It is hence of crucial importance to understand the qualities that can be expected from practically implementable transformations. Here, we consider the limitations on charging batteries when restricting to the feasibly realizable family of Gaussian unitaries. We derive optimal protocols for general unitary operations as well as for the restriction to easier implementable Gaussian unitaries. We find that practical Gaussian battery charging, while performing significantly less well than is possible in principle, still offers asymptotically vanishing relative charge variances and fluctuations.
\end{abstract}

\maketitle


\section{Introduction}

Quantum thermodynamics (QT) deals with the manipulation and transfer of energy and entropy at the quantum scale. How well one can transfer energy depends greatly on the information one has about a system~\cite{GooldHuberRieraDelRioSkrzypczyk2016, MillenXuereb2016, VinjanampathyAnders2016}. Consequently, {the system entropy quantifying this information} is rendered an important quantity for achievable state transformations~\cite{BrandaoHorodeckiNgOppenheimWehner2015}. At fixed energy, the entropy is maximized for thermal states, which allows for the definition of thermal equilibrium characterized by the emergent notion of temperature. A system in such a thermal equilibrium with an environment at temperature $T$ is thermodynamically useless in the sense that its energy cannot be extracted as work~\cite{BrandaoHorodeckiOppenheimRenesSpekkens2013, Mueller2017}. Therefore, much effort has been invested into understanding the emergence of equilibration and thermalization in quantum systems~\cite{GogolinEisert2016}. {At the same time, quantifying extractable energy and identifying achievable transformations crucially depends on} the control one assumes to have about microscopic degrees of freedom. {For instance}, acting only upon individual quantum systems {from whom work is to be extracted} gives rise to the notion of passive states~\cite{PuszWoronowicz1978} which cannot yield any work in cyclic Hamiltonian processes, even if the entropy at a given energy is far below the thermal entropy~\cite{PerarnauLlobetHovhannisyanHuberSkrzypczykTuraAcin2015}. However, even non-passive states may still require complex operations and precise control over large Hilbert spaces that make them practically unfeasible sources of work. A recent focus of thermodynamic resource theories has thus been to investigate the role of precise control and practically implementable operations for achieving desired work extraction~\cite{BrownFriisHuber2016, PerryCwiklinskiAndersHorodeckiOppenheim2015, LostaglioAlhambraPerry2017, MazurekHorodecki2017} {and refrigeration}~\cite{ClivazSilvaHaackBohrBraskBrunnerHuber2017}.

{The resource-theoretic view on quantum thermodynamics of course extends beyond the task of work extraction, and generally aims to identify the ultimate limitations of all single-shot processes}~\cite{BrandaoHorodeckiOppenheimRenesSpekkens2013, HorodeckiOppenheim2013b, GourMuellerNarasimhacharSpekkensYungerHalpern2015}{. More specifically, viewing quantum thermodynamics as a resource theory entails either the ability to perform any unitary operation induced by a Hamiltonian $H(t)$ with control parameter $t$ (i.e., the case of ``driven" or controlled operations), or applying arbitrary ``thermal operations", i.e., global energy conserving operations on the chosen system and arbitrary many auxiliary systems. When these auxiliary systems can feature coherence w.r.t. the energy eigenbasis (i.e., if coherent ``batteries" are provided), these paradigms become equivalent}~\cite{Aberg2014, MalabarbaShortKammerlander2015}{. However, neither paradigm limits the complexity of the allowed operations, requiring arbitrary coherent energy shifts in subsystems from which energy is extracted or in which energy is stored}~\cite{SkrzypczykShortPopescu2013, SkrzypczykShortPopescu2014}{. This can lead to genuine quantum advantages, e.g., for the charging power of $N$-qubit batteries}~\cite{BinderVinjanampathyModiGoold2015, CampaioliPollockBinderCeleriGooldVinjanampathyModi2017} {and for small, finite-dimensional systems (e.g., few-qubit registers) such full control over the quantum systems may reasonably be expected. However, for larger systems such as registers of many qubits, arbitrary global operations may be difficult to realize and call for more specialized practical solutions}~\cite{FerraroCampisiAndolinaPellegriniPolini2018}{. In particular, this applies to infinite-dimensional quantum systems such as (ensembles of) harmonic oscillators. Besides the paradigmatic two-dimensional Hilbert spaces of qubits often favoured in information-theoretic approaches to quantum mechanics, harmonic oscillators play a crucial role for the description of physical systems in quantum optics and quantum field theory. Indeed, all current realistic proposals for and implementations of quantum machines involve at least one Hilbert space corresponding to a harmonic oscillator. Examples for such systems include superconducting resonators}~\cite{HoferSouquetClerk2016, HoferPerarnauLlobetBohrBraskSilvaHuberBrunner2016}{, modes of the electromagnetic field in a cavity}~\cite{MitchisonHuberPriorWoodsPlenio2016}{, or vibrational modes of trapped ions}~\cite{MaslennikovDingHablutzelGanRouletNimmrichterDaiScaraniMatsukevich2017, RossnagelDawkinsTolazziAbahLutzSchmidtKalerSinger2016}. {It is hence of conceptual significance to understand the fundamental as well as the practical limitations for thermodynamic tasks in such infinite-dimensional continuous-variable (CV) systems. In particular, full control over such systems automatically implies the ability to create coherence between energy levels with arbitrarily large separation.}

{In contrast, a class of operations that can typically be realized comparatively simply in quantum optical realizations of CV systems is that of Gaussian unitaries}~\cite{Weedbrooketal2012}{. In the context of driven quantum systems these operations naturally appear as most straightforwardly implementable in the hierarchy of driving Hamiltonians since they require $H(t)$ to be at most quadratic in the system's creation and annihilation operators. Indeed, most natural interactions are appropriately described by such ``bipartite terms" (usually resonant energy exchanges), whereas the creation of higher order terms is a challenge that is usually addressed only in a perturbative way. For the task of work extraction, the restriction of thermodynamic operations on CV systems to Gaussian transformations brings about the notion of \emph{Gaussian passivity}}~\cite{BrownFriisHuber2016}{, which encompasses states that are potentially non-passive, but are passive w.r.t. Gaussian transformations.} Once work has been extracted, one would of course also like to put it to use, potentially at a later time. This requires the previously obtained energy to be stored and distributed. It is hence expected that practical limitations applying to work extraction \textemdash in particular, the restriction to Gaussian operations \textemdash will also be relevant for these tasks.

\begin{figure}[ht!]
\label{fig:charging scheme}
\begin{center}
\includegraphics[width=0.49\textwidth,trim={0cm 0mm 0cm 0mm}]{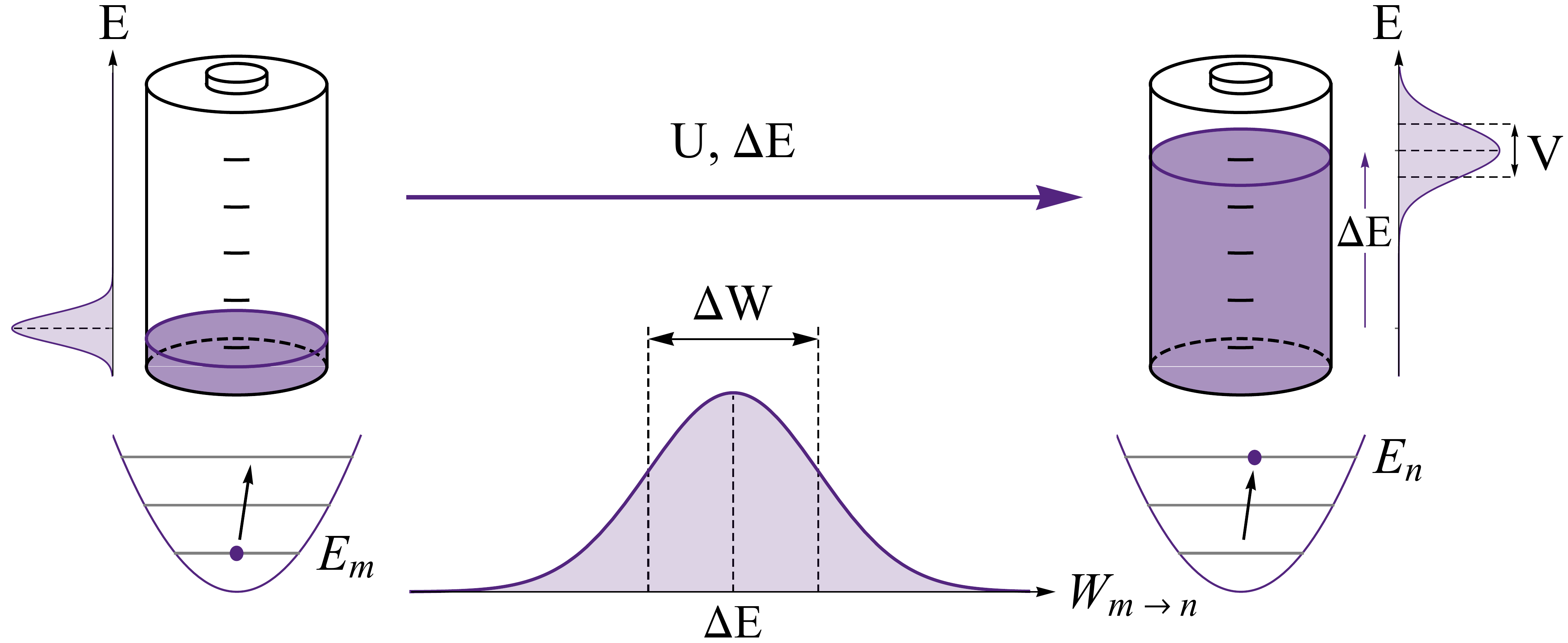}
\end{center}
\caption{
\textbf{Quantum battery charging:} The average energy of an initially thermal battery is unitarily increased by $\Delta E$. The fluctuations of the final charge and of the energy supply can be quantified by the variance $V$ of the energy distribution in the final battery, and by the average square deviation $(\Delta W)^{2}$ of the transitions $W_{m\rightarrow n}=E_{m}-E_{n}$ between the levels $m$ and $n$ from the average energy supply $\Delta E$.}
\end{figure}

Following the full characterization of Gaussian passivity~\cite{BrownFriisHuber2016}, we hence aim to quantify the limitations imposed by the restriction to Gaussian unitaries on the task of energy transfer to suitable quantum optical storage devices, i.e., charging batteries. More precisely, we consider ensembles of harmonic oscillators as batteries. These batteries are assumed to be initially uncharged in the sense that they contain no extractable work. That is, we consider the empty batteries to be in thermal equilibrium with the environment, and describe them by thermal states at the ambient temperature. We then study the task of unitarily increasing their energy by a fixed increment $\Delta E$. Although such unitaries always exist in infinite-dimensional Hilbert spaces, unitaries achieving a given energy increase are not uniquely determined by $\Delta E$, and may offer different charging precision, speed, and energy fluctuations during the charging process.

Here, we focus on two quantities characterizing the reliability of the charging process, as illustrated in Fig.~\ref{fig:charging scheme}. First, the charging precision, represented by the energy variance $V$ of the battery, which is of interest since it is desirable that a charged battery is able to deliver the expected energy, not hazardously much more energy or disappointingly much less. 
Second, we consider the fluctuations during the charging process, captured by $(\Delta W)^{2}$ the average square deviation of the energy transitions from the average energy supply. {While the variance quantifies the usefulness of the battery in terms of potential fluctuations occurring when discharging the loaded battery, the variance loosely speaking only captures half of the problem. That is, taking into account the initial distribution of energies one may also be interested in the energy fluctuations during the charging process. The resulting distribution is often called ``fluctuating work"}~\cite{CampisiHaenggiTalkner2011, AlhambraMasanesOppenheimPerry2016, RichensMasanes2016} {and characterizes the distribution of work if one were to measure the battery in the energy eigenbasis at the beginning and end of the charging protocol.} For both of these characteristics we determine the ultimate limitations during arbitrary unitary charging processes by designing optimal protocols. We then specialize to Gaussian unitaries, for which we identify the optimal and worst charging protocols. In comparison, we find that Gaussian unitaries perform significantly less well than is possible in principle. Nonetheless, Gaussian battery charging can asymptotically achieve vanishing relative fluctuations $V/\Delta E$ and $(\Delta W)^{2}/\Delta E$ for large input energies by way of simple combinations of displacements and single-mode squeezing. Our results hence provide insights into both the fundamental and practical limitations of charging quantum optical batteries.

This article {is} structured as follows. In Sec.~\ref{sec:framework}, we set the stage for the investigation and define the quantities of interest. We then present an investigation of the fundamental limitation of charging quantum batteries using arbitrary unitaries in Sec.~\ref{sec:Single-mode battery in the ground state}, before we restrict to Gaussian transformations in Sec.~\ref{sec:battery charging with Gaussian unitaries}. Finally, we draw conclusions in Sec.~\ref{sec:discussion}.


\section{Charging a quantum battery}\label{sec:framework}
\vspace*{-1mm}



As battery systems to be charged we consider a number of bosonic modes (i.e., an ensemble of harmonic oscillators) initially in thermal states $\tau(\beta)=\exp(-\beta H)/\mathcal{Z}$, where $\mathcal{Z}=\tr\bigl(\exp(-\beta H)\bigr)$ is the partition function, $\beta=1/T$ is the inverse temperature of the battery (we use units where $\hbar=k_{\protect\raisebox{-0pt}{\tiny{B}}}=1$ throughout), and {$H=\sum_{j}\omega_{j}a_{j}^{\dagger}a_{j}$} is the system Hamiltonian. The mode operators {$a_{j}$} and {$a_{j}^{\dagger}$} satisfy the usual commutation relations {$\comm{a_{j}}{a_{k}^{\dagger}}=\delta_{jk}$} and {$\comm{a_{j}}{a_{k}}=0$}. For such a non-interacting Hamiltonian, the initial state is a product state $\tau(\beta)=\bigotimes_{i}\tau_{i}(\beta)$. The single-mode Gibbs states $\tau_{i}(\beta)$ can be written as
\begin{align}
    \tau_{i}(\beta) &=\,(1-e^{-\beta\omega_{i}})\sum_{n}\,e^{-n\beta\omega_{i}}\,\ket{n_{i}}\!\!\bra{n_{i}}\,
    \label{eq:single mode thermal states}
\end{align}
with respect to their respective Fock bases $\{\ket{n_{i}}\}$, where $a_{i}\ket{n_{i}}=\sqrt{n_{i}}\ket{(n-1)_{i}}$ and $a_{i}^{\dagger}\ket{n_{i}}=\sqrt{(n+1)_{i}}\ket{(n+1)_{i}}$. This choice of initial state ensures that the batteries are truly empty at first, i.e., the initial state is passive for any number of such batteries because the Gibbs state is completely passive (uniquely at fixed energy).

We are then interested in applying a unitary transformation $U$ to raise the average energy by $\Delta E$, transforming the initial state $\tau(\beta)$ to a final state $\rho=U\tau U^{\dagger}$, i.e.,
\begin{align}
    \Delta E    &=\,E(\rho)-E\bigl(\tau(\beta)\bigr)\,=\,\tr\bigl(H[\rho-\tau(\beta)]\bigr)\,\nonumber\\[1mm]
    &=\,\tr(H\rho)-\sum\limits_{n}\frac{\omega_{n}}{e^{\beta\omega_{n}}-1}.
\end{align}
We quantify the charging precision via the increase of the standard deviation of the system Hamiltonian, that is, one of the quantities that we are interested in is
\begin{align}
    \Delta\sigma    &=\,\sqrt{V(\rho)}-\sqrt{V(\tau)},
\end{align}
where the variance w.r.t. $H$ is given by
\begin{align}
    V(\rho) &=\,\bigl(\Delta H(\rho)\bigr)^{2}\,=\,\tr(H^{2}\rho)-\bigl(\tr(H\rho)\bigr)^{2}.
\end{align}

Besides the precision of the final battery charge, one may also care about other quantities, for instance, the energy fluctuations\footnote{Note that we use a definition for energy fluctuations suitable for the task at hand, which differs from fluctuations in the sense of thermodynamical fluctuation relations~\protect\cite{EspositoHarbolaMukamel2009}.} of the charging process. That is, we consider the average squared deviation from the average energy increase, given by
\begin{align}
    (\Delta W)^{2} &=\,\sum\limits_{m,n}p_{m\rightarrow n}(W_{m\rightarrow n}-\Delta E)^{2}\,,
    \label{eq:work fluctuations}
\end{align}
where $W_{m\rightarrow n}=E_{n}-E_{m}$ is the work relating the $m$-th and $n$-th energy levels, with $H\ket{n}=E_{n}\ket{n}$, and
\begin{align}
    p_{m\rightarrow n}  &=\,p_{m}\,|\bra{n}U\ket{m}|^{2}
\end{align}
is the probability of a transition from the $m$-th to the $n$-th energy eigenstate starting from the initial state $\tau(\beta)$ with diagonal elements $p_{n}=\bra{n}\tau\ket{n}$. To better understand the quantity $\Delta W$ it is useful to note that we can write the squared work fluctuation as the variance of the operator $H_{\Delta}=\tilde{H}-H$ in the thermal state, where $\tilde{H}=U^{\dagger}HU$, i.e.,
\begin{align}
    (\Delta W)^{2} &=\,
    \expval{H_{\Delta}^{2}}_{\tau}-\expval{H_{\Delta}}_{\tau}^{2}
    \nonumber\\[1mm]
     &=\,
     (\Delta\tilde{H}_{\tau})^{2}
     +(\Delta H_{\tau})^{2}
     \text{{-}}2\operatorname{Cov}(\tilde{H},H)\,,
     \label{eq:fluctuations 1}
\end{align}
where the covariance is given by
\begin{align}
    \operatorname{Cov}(\tilde{H},H) &=\,\tfrac{1}{2}\expval{\{\tilde{H},H\}_{+}}-\expval{\tilde{H}}\expval{H},
    \label{eq:covariance general}
\end{align}
and $\{\tilde{H},H\}_{+}=\tilde{H}H+H\tilde{H}$ denotes the anticommutator. In general, the operators $\tilde{H}$ and $H$ need not commute, but since the initial thermal state is diagonal in the energy eigenbasis, we can further simplify Eq.~(\ref{eq:fluctuations 1}) and obtain
\begin{align}
    (\Delta W)^{2} &=\,V(\rho)+V(\tau)
    -2\,\bigl[\tr(\tilde{H}H \tau)-E(\tau)E(\rho)\bigr].
    \label{eq:squared work fluctuation}
\end{align}
The squared increase of the standard deviation, in comparison, can be written as
\begin{align}
    (\Delta \sigma)^{2} &=\,V(\rho)+V(\tau)-2\sqrt{V(\rho)V(\tau)}\,.
\end{align}
Since one can write $E(\tau)E(\rho)=\tr(U^{\dagger}H U \expval{H}_{\tau}\tau)$, it is easy to see the charging precision and fluctuations coincide when the initial state is an eigenstate of the Hamiltonian, because $\expval{H}_{\tau}\tau=E_{n}\ket{n}\!\!\bra{n}=H\tau$ and $V(\tau)=0$. In this case (which, in our scenario only occurs for the ground state since our initial state is a thermal state), one has $(\Delta W)^{2}=(\Delta\sigma)^{2}=V(\rho)$.


\section{Fundamental Limits for Battery Charging}\label{sec:Single-mode battery in the ground state}
\vspace*{-1mm}

In this section, we investigate the fundamental limits on the charging precision and fluctuations. As we shall see, optimal protocols can be constructed that minimize either the variance of the final energy or the fluctuations during the charging process, but these do not coincide for finite temperatures. However, the involved operations are often rather complicated in the sense that they require very specific interventions in particular subspaces of the infinite-dimensional Hilbert space, tailored to the initial temperature and energy supply. The results obtained in this section hence illustrate what is in principle possible and provide a benchmark for the precision and fluctuations achievable with Gaussian unitaries.


\subsection{Fundamental limits for zero temperature}\label{sec:Fundamental limitations ground state single mode}
\vspace*{-2mm}

Let us first consider a simple example to set the stage for a further, in-depth investigation. To this end, we consider a single-mode battery that is initially in the ground state, i.e., $H=\omega a^{\dagger}a$ and $\tau=\ket{0}\!\bra{0}$. In this case, the work fluctuations and charging precision coincide and are given by
\begin{align}
      (\Delta\sigma)^{2} &=(\Delta W)^{2}=(\Delta H(\rho))^{2},
\end{align}
and $\rho=U\ket{0}\!\bra{0}U^{\dagger}=\ket{\psi}\!\bra{\psi}$ is a pure state. Since the Hilbert space in question is infinite-dimensional the energy variance of the final state is not bounded from above. This can be seen by choosing a superposition of the form $\ket{\psi}=\sqrt{q}\ket{0}+\sqrt{1-q}\ket{k}$ with $k=(1\!-\!q)^{-1}\Delta\epsilon$, such that $\tfrac{\expval{H}_{\psi}}{\omega}=\tfrac{\Delta E}{\omega}\equiv\Delta\epsilon$. A simple calculation then reveals that
\begin{figure}[ht!]
\label{fig:single mode no restrictions}
\begin{center}
\includegraphics[width=0.45\textwidth,trim={0cm 0mm 0cm 0mm}]{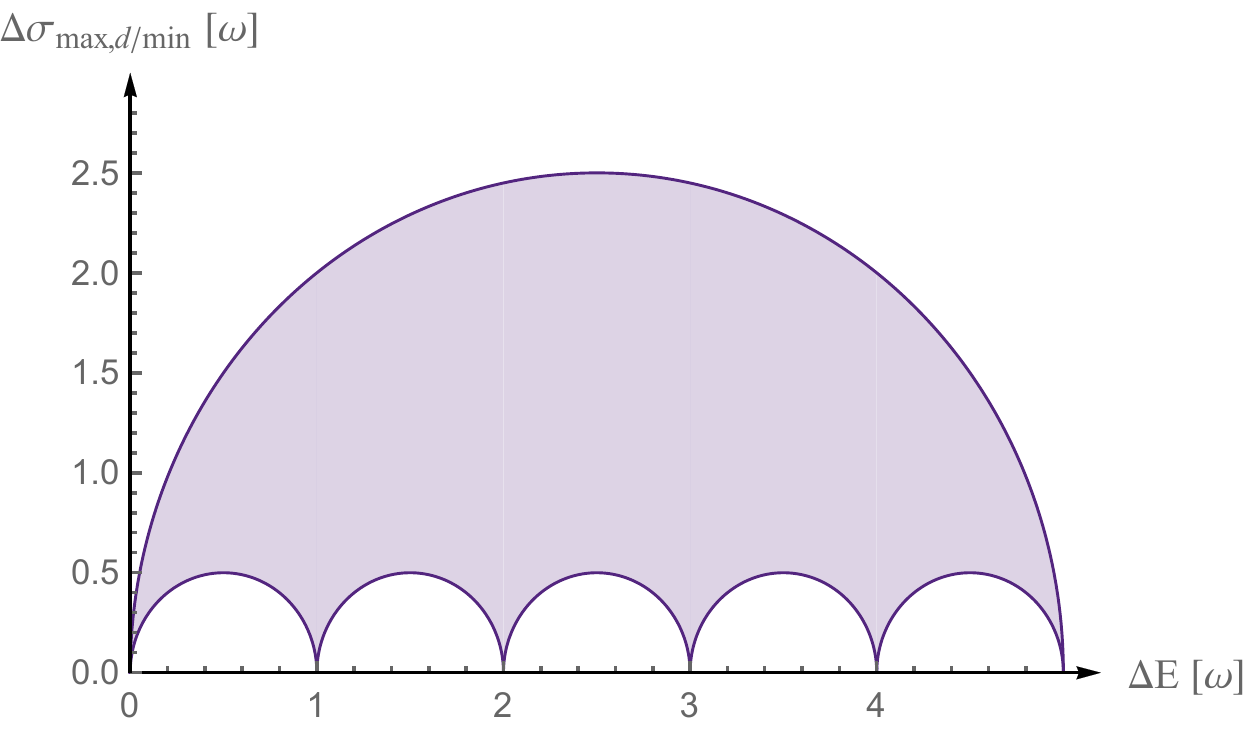}
\end{center}
\caption{
\textbf{Unrestricted battery charging:} The maximal and minimal variances of the energy that are in principle possible for a battery that starts in its ground state and is being charged by $\Delta E$ are shown (in units of $\omega$, with $\hbar=1$) for a system of dimension $d=6$. When the Hilbert space is infinite-dimensional, the lower bound periodically repeats, but the upper bound is no longer finite for any value of $\Delta E$. Since the initial temperature vanishes, the bounds shown also apply to the charging fluctuation $\Delta W$.}
\end{figure}
\noindent
\begin{align}
    \left(\tfrac{\Delta\sigma}{\omega}\right)^{2}    &=\,\Delta\epsilon(k-\Delta\epsilon).
\end{align}
In other words, for any chosen energy $\Delta E$ one can make $k$ (and hence $\Delta \sigma=\Delta W$) arbitrarily large by simultaneously choosing $q$ sufficiently close to $1$. So for arbitrarily small energies, the energy variance and the fluctuations during the charging process may increase by an arbitrary amount. However, it is also clear that this is an artefact of the infinite-dimensional character of the system. If the dimension $d$ of the system is finite (or there is some cutoff energy), then the maximal variance is obtained for a superposition of the eigenstates $\ket{0}$ and $\ket{d-1}$, with minimal and maximal eigenvalues, respectively, resulting in
\begin{align}
    \tfrac{(\Delta\sigma_{\mathrm{max,d}})^{2}}{\omega^{2}}  &=\,\Delta\epsilon\bigl((d-1)-\Delta\epsilon\bigr)\,.
\end{align}
The minimal achievable variance for any given energy is obtained by unitarily rotating to a superposition of the two energy eigenstates $\ket{n}$ and $\ket{n+1}$ that are closest to the available energy, i.e., such that $n\leq \Delta\epsilon\leq n+1$. More specifically, we have $\ket{\psi}=U\ket{0}=\sqrt{p}\ket{n}+\sqrt{1-p}\ket{n+1}$ with
\begin{align}
    p   &=\,\lceil \Delta\epsilon \rceil\,-\,\Delta\epsilon\,,
\end{align}
resulting, after some algebra, in a variance of
\begin{align}
    \left(\tfrac{\Delta\sigma_{\mathrm{min}}}{\omega}\right)^{2}  &=
     \bigl(\Delta\epsilon-\lfloor\Delta\epsilon\rfloor\bigr)\bigl(\lceil\Delta\epsilon\rceil-\Delta\epsilon\bigr).
     \label{eq:zero temp opt var}
\end{align}
Crucially, $\Delta\sigma_{\mathrm{min}}=0$ whenever $\Delta E$ is an integer multiple of the oscillator frequency, and the maximal value of $\Delta\sigma_{\mathrm{min}}$ is $\tfrac{\omega}{2}$, as illustrated in Fig.~\ref{fig:single mode no restrictions}.


\subsection{Fundamental precision limits for arbitrary temperatures}\label{sec:Fundamental limitations thermal states single mode}


Having understood the simple case of optimally charging a battery initially in the ground state, we now want to move on to the case of thermal battery states. On the one hand, the worst-case scenario immediately carries over from the situation discussed in the previous section. That is, in an infinite-dimensional system one may always find a unitary transformation that increases the energy by an arbitrarily small amount, while increasing the variance arbitrarily strongly. This can be seen by just noting that the two-level rotation used to rotate between the ground state and the level $\ket{k}$ can also be applied to thermal states. The only difference is that the corresponding probability weights are now different from $1$ and $0$ initially.  In contrast, the upper bound for the variance in a finite-dimensional system is always finite.

The optimally achievable charging precision, on the other hand, requires a more intricate analysis. The task at hand is to specify the state of minimal energy variance $V(\rho)$ at a fixed average energy within the unitary orbit of a thermal state at a given temperature. In general, we cannot give a closed expression relating this minimal variance to the energy input and the initial temperature. However, one may formulate a protocol that provides (one of) these minimal variance states. Here, we will give a short, intuitive description of this protocol, and provide a detailed step-by-step account in Appendix~\ref{sec:optimal general charging protocol}.

Let us now briefly explain the working principle of the optimal-precision charging protocol. First, 
recall that the initial thermal state has a density operator $\tau(\beta)$ that is diagonal in the energy eigenbasis with probability weights $p_{n}$ decreasing with increasing energies $E_{n}$. The average energy $E\bigl(\tau(\beta)\bigr)$ is determined by the initial temperature and we hence know the target energy $E(\rho)=E(\tau)+\Delta E$ for any energy input. We can then naively apply two-level rotations to reorder the probability weights on the diagonal such that the largest weight $p_{0}$ is shifted to the eigenstate whose energy is closest to the target energy, the second largest weight is shifted to the second-closest eigenstate to $E(\rho)$, and so on. This procedure results in the unique state $\tilde{\rho}(\beta)$ within the unitary orbit of $\tau(\beta)$ whose average squared deviation $\tilde{V}$ from the target energy is minimal.

\begin{figure}[ht!]
\label{fig:optimal charging}
\begin{center}
(a)\includegraphics[width=0.47\textwidth,trim={0cm 0mm 0cm 0mm}]{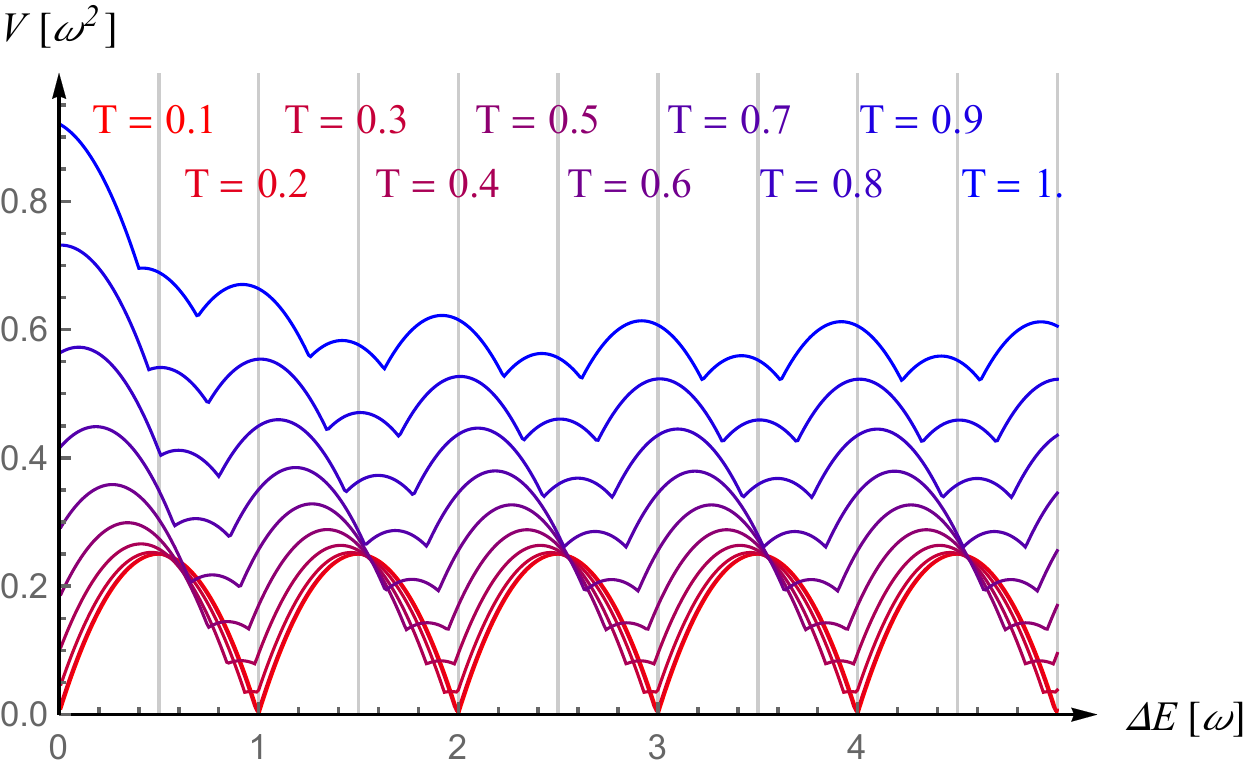}
(b)\includegraphics[width=0.47\textwidth,trim={0cm 0mm 0cm 0mm}]{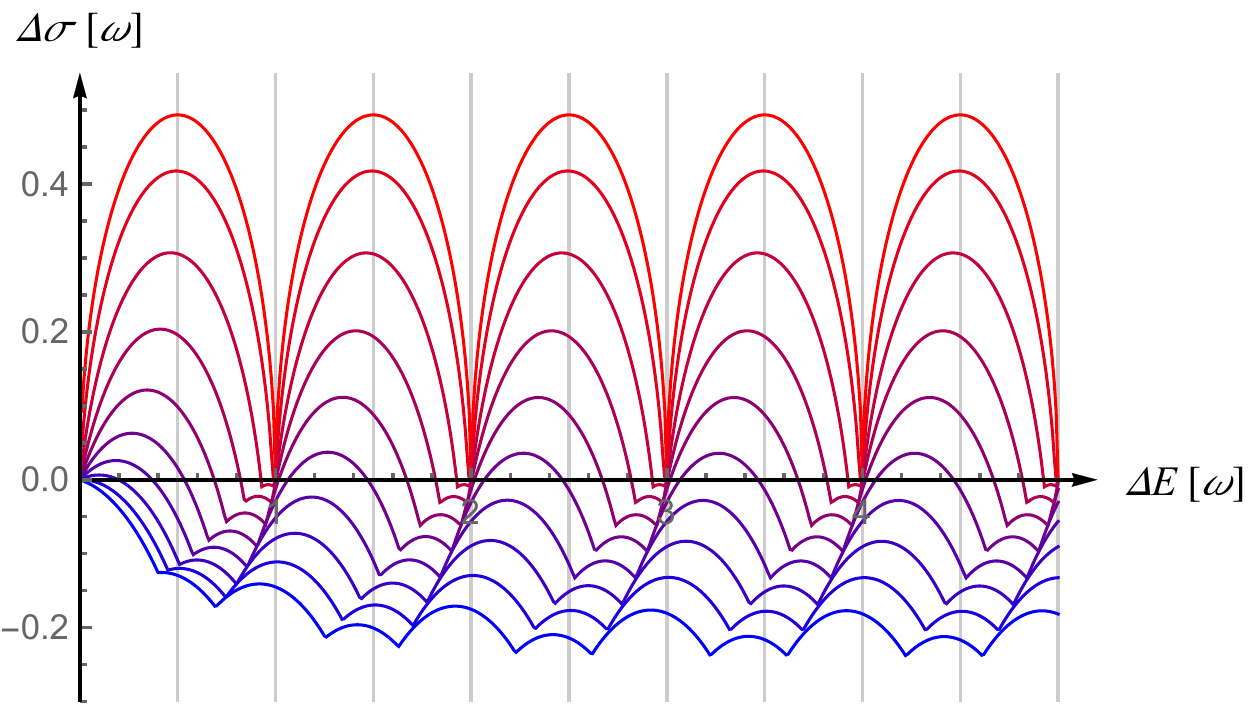}
\end{center}
\caption{
\textbf{Optimal precision charging of thermal battery:} The minimal variance $V(\rho)$ (in units of $\omega^{2}$) and the optimal standard deviation change $\Delta\sigma$ (in units of $\omega$) that are in principle achievable for charging a quantum battery at initial temperature $T=0.1$ to $T=1$ (in steps of $0.1$ and units of $\omega$, bottom to top in (a) and top to bottom in (b)) are plotted against the energy input $\Delta E/\omega$ in (a) and (b), respectively.
}
\end{figure}

Unfortunately, this state {does not} generally have the desired target average energy, i.e., $\tilde{E}=E(\tilde{\rho})\neq E(\rho)$. Consequently, the average squared deviation from $E(\rho)$ is generally not equal to the energy-variance, $\tilde{V}\neq V$. Moreover, both the cases $\tilde{E}>E$ and $\tilde{E}<E$ can occur and one therefore has to adjust the energy accordingly. This can be done by sequences of two-level rotations that change the energy by $\Delta\tilde{E}$ and increase the average squared deviation from $E$ by $\Delta\tilde{V}$. An ordering of these operations that is optimal is obtained when performing them in the order of increasing values of $\Delta\tilde{V}/|\Delta\tilde{E}|$, starting with the smallest, i.e., when the increase of $\tilde{V}$ per unit energy change is as small as possible. One carries on with this protocol until the desired target energy is reached, in which case the final value of $\tilde{V}$ becomes the variance of the energy $V$. The resulting variances for given energy input for {a harmonic oscillator} are illustrated in Fig.~\ref{fig:optimal charging}. It can be seen that for higher initial temperatures this optimal protocol can lead to decreasing variances in the battery state. {Also note that the working principle of this optimal protocol is unchanged if one considers a finite-dimensional system instead and differences only arise because of the finite maximal energy input at any given temperature.}


\subsection{Fundamental precision limits for multi-mode batteries}\label{sec:Fundamental limitations multi mode}


After obtaining the fundamental limits on the precision of charging a single-mode battery, it is of course natural to ask which possibilities arise when several such batteries are available. The worst case scenario for multiple modes trivially translates from our previous analysis. Since the variance for any given energy input is not bounded from above for single-mode batteries, the same is also true for many modes.

To understand what can be achieved in the best case for multiple batteries, let us first consider the two-mode case, i.e., two batteries labelled $A$ and $B$ that are initially in a thermal state $\tau\subtiny{0}{0}{A}\otimes\tau\subtiny{0}{0}{B}$. We are now interested in an increase of the average energy $E(\rho)=\tr\bigl(\rho H\bigr)$ w.r.t. $E(\tau)$, where the bipartite Hamiltonian is $H=H\subtiny{0}{0}{A}+H\subtiny{0}{0}{B}=\omega\subtiny{0}{0}{A}a\subtiny{0}{0}{A}^{\dagger}a\subtiny{0}{0}{A}+\omega\subtiny{0}{0}{B}a\subtiny{0}{0}{B}^{\dagger}a\subtiny{0}{0}{B}$. The energy variance is then given by
\begin{align}
    (\Delta H)^{2}  &=(\Delta H\subtiny{0}{0}{A})^{2}+(\Delta H\subtiny{0}{0}{B})^{2}+2\operatorname{Cov}(H\subtiny{0}{0}{A},H\subtiny{0}{0}{B}),
\end{align}
where the covariance of Eq.~(\ref{eq:covariance general}) for the local (and hence commuting) operators $H\subtiny{0}{0}{A}$ and $H\subtiny{0}{0}{B}$ is
\begin{align}
    \operatorname{Cov}(H\subtiny{0}{0}{A},H\subtiny{0}{0}{B}) &=\,\expval{H\subtiny{0}{0}{A}\otimes H\subtiny{0}{0}{B}}-\expval{H\subtiny{0}{0}{A}}\expval{H\subtiny{0}{0}{B}}.
    \label{eq:covariance}
\end{align}
\begin{figure}[ht!]
\label{fig:optimal charging two modes}
\begin{center}
\includegraphics[width=0.47\textwidth,trim={0cm 0mm 0cm 0mm}]{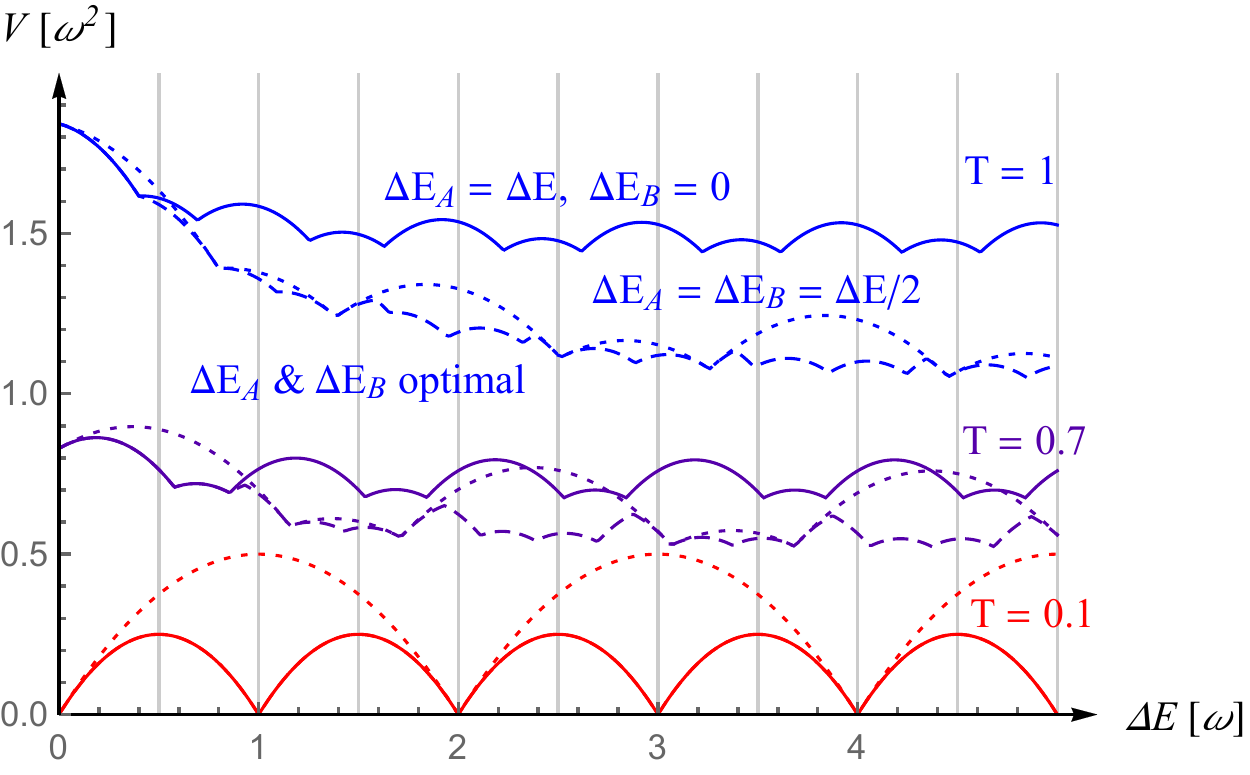}
\vspace*{-2mm}
\end{center}
\caption{\textbf{Precision improvement for two-mode batteries:} {The charging precision in terms of the overall variance $V(\rho)$ (in units of $\omega^{2}$) is shown for a battery consisting of two modes with equal frequencies $\omega$ for sample temperatures of $T=0.1$ (red, bottom), $T=0.7$ (purple, middle) to $T=1$ (blue, top) in units of $\omega$. For each temperature, three curves are shown corresponding to different local unitary charging protocols pertaining to different distribution of the overall energy input $\Delta E$ into the energy increases $\Delta E_{A}$ and $\Delta E_{B}$ of the two modes labelled $A$ and $B$, respectively. The solid curves indicate that all energy is stored in one of the modes only, $\Delta E_{1}=\Delta E$, $\Delta E_{2}=0$. Dotted lines correspond to equal charging energies for both modes, $\Delta E_{A}=\Delta E_{B}=\Delta E/2$, and dashed lines represent optimally splitting the charge between both modes. For the sake of numerical optimization we have chosen integer multiples of $\omega/20$ as indivisible units of energy charge, meaning optimality here means the optimal choice of $m,n\in\mathbb{N}_{0}$ such that $\Delta E_{A}=m\tfrac{\omega}{20}$, $\Delta E_{B}=n\tfrac{\omega}{20}$, and $(m+n)\tfrac{\omega}{20}=\Delta E$. Note that for the lowest temperature shown ($T=0.1$), the solid and dashed lines are virtually indistinguishable, meaning that there is no (distinguishable) advantage in splitting the energy between the modes. However, such an advantage is clearly visible for higher temperatures.}
}
\end{figure}
\noindent
For a local unitary charging protocol, i.e., where $U=U\subtiny{0}{0}{A}\otimes U\subtiny{0}{0}{B}$, the initial thermal states remain uncorrelated and the covariance vanishes. That is, the final state $\rho\subtiny{0}{0}{AB}$ is a product state $\rho\subtiny{0}{0}{AB}=\rho\subtiny{0}{0}{A}\otimes\rho\subtiny{0}{0}{B}$. In such a case not only the average energies but also the variances are additive. Inspection of Fig.~\ref{fig:optimal charging} then shows that having two or more batteries available can be beneficial even when they are charged independently. For instance, when the supplied energy would lead to a local maximum of the variance if all energy is stored in one battery, it may be prudent to reduce the energy supply to this battery to reach a (local) minimum instead. The remnant energy can then be stored in a second battery. When the two modes have the same frequency and the initial temperature is nonzero the resulting overall variance is then smaller {than or equal to} that of charging only one battery, as we can see from {Fig.}~\ref{fig:optimal charging two modes}. In short, the availability of several {battery modes} at potentially different frequencies hence provides a certain flexibility to reach local minima of the variances of the individual batteries{, but the exact performance for a given set of battery modes requires to be worked out on a case-by-case basis.}

For unitaries that are not local and can correlate the two batteries, the situation is {even} more involved but {in principle} such unitaries may help to achieve an even better performance. To see this, let us return to the optimal protocol of the last section. In the first step of this protocol, the probability weights of the initial thermal state are reordered to create a distribution that is as narrow as possible around the target energy. The resulting state is diagonal in the energy eigenbasis. Since this is a product basis {w.r.t the tensor product structure of different modes}, the state is still uncorrelated. However, in the second step, where the energy of the distribution is adjusted to the target energy, two-level rotations with optimal ratios $\Delta\tilde{V}/|\Delta\tilde{E}|$ may occur between states $\ket{m,n}$ and $\ket{m\pr,n\pr}$ with $m\neq m\pr$ and $n\neq n\pr$ and hence correlate the systems. For batteries at different frequencies there can thus be an advantage in introducing (specific) correlations, whereas a situation as just described can always be avoided for batteries with equal frequencies. In the general case of arbitrary frequencies it is interesting to note though, that the creation of correlations may be marginally helpful but is not the key ingredient. This is in contrast to recent results on the charging power, where {the ability to create quantum correlations, i.e., access to entangling operations (albeit not necessarily the actual creation of entanglement)} can be extremely useful~\cite{BinderVinjanampathyModiGoold2015, CampaioliPollockBinderCeleriGooldVinjanampathyModi2017}.

To reach optimality it nonetheless remains to be determined how the energy can be optimally split between the oscillators, or invested in correlations. Unfortunately, this is difficult to answer in general, and is even rather complicated for uncorrelated charging due to the non-monotonic behaviour of the optimal single-mode charging protocol illustrated in Fig.~\ref{fig:optimal charging}{, which is illustrated in Fig.}~\ref{fig:optimal charging two modes}. There, the specific optimal splitting depends on the initial temperature, the specific energy input, and the (number and) frequencies of the {battery modes} involved. The optimal performance hence has to be determined on a case-by-case basis. However, one can state quite generally that the optimal final variance {of the joint system} is never larger than the optimal variance when all the energy is stored only in one of the {modes}. In other words, having several {battery modes} available is never detrimental. Indeed, having more empty batteries at different frequencies at one's disposal can be considered a nontrivial resource for precise charging.

Having discussed which charging precisions can be achieved in principle, let us briefly turn to the fundamental limitations arising for the charging fluctuations.


\subsection{Fundamental fluctuation limits for arbitrary temperatures}\label{sec:Fundamental limitations fluctuations thermal states single mode}


To complete the investigation of the fundamental restrictions of charging a quantum battery, let us consider a protocol that minimizes the fluctuations $\Delta W$. For simplicity, let us start with the case where the input energy is exactly one unit, $\Delta\epsilon=1$. Then the infinite-dimensional Hilbert space allows keeping the fluctuations arbitrarily small. To achieve this, we perform a unitary permutation operation on the first $N$ energy levels that shifts the weight $p_{n}=(1-e^{-\beta\omega})e^{-n\beta\omega}$ from the level $n$ to the level $n+1$ for $n=0,\ldots,N-1$, while the last weight $p_{N}$ is shifted to the ground state level. In the limit $N\rightarrow\infty$, the energy is increased by $\Delta E=\omega$ and since $\bra{m}U\ket{n}=\delta_{m,n+1}$, the fluctuations vanish.

When the input energy is less than one unit, i.e., when $0<\Delta\epsilon<1$, the fluctuations do not vanish, but can be minimized in a simple way. Suppose that we perform the same permutation as before, but start shifting weights upwards at some finite $n=k$ rather than at $n=0$, such that a vanishingly small weight is placed on the $k$-th level. The corresponding final state energy (in units of $\omega$) would be
\begin{align}
    \tilde{\epsilon}  &=\epsilon_{0}+\tilde{\Delta\epsilon}=\!\!\sum\limits_{n=0}^{k-1}n\,p_{n}\!+\!\!\!\!\!\sum\limits_{n=k+1}^{\infty}\!\!\!n\,p_{n-1}
    \!=\!\!\sum\limits_{n=0}^{\infty}n\,p_{n}\!+\!\!\sum\limits_{n=k}^{\infty}p_{n},
\end{align}
where $\epsilon_{0}=\sum_{n=0}^{\infty}n p_{n}=E(\tau)/\omega$. The increase w.r.t. the initial state would hence be
\begin{align}
    \tilde{\Delta\epsilon}  &=\!\sum\limits_{n=k}^{\infty}p_{n}=\!\sum\limits_{n=k}^{\infty}(1-e^{-\beta\omega})e^{-n\beta\omega}\,=\,e^{-k\beta\omega}.
\end{align}
Now, generally, $(\beta\omega)^{-1}\ln(1/\Delta\epsilon)$ is not an integer and, consequently, the energy shift upwards starting from $k=\tilde{k}:=\lceil(\beta\omega)^{-1}\ln(1/\Delta\epsilon)\rceil$ is not enough, $\Delta\epsilon_{\mathrm{I}}:=e^{-\tilde{k}\beta\omega}\leq\Delta\epsilon$. However, if we perform the shift from $\tilde{k}$ onwards nonetheless, the difference $\Delta\epsilon_{\mathrm{I\hspace*{-0.5pt}I}}=\Delta\epsilon-\Delta\epsilon_{\mathrm{I}}$ can be obtained by continuously rotating between the level $\tilde{k}-1$ and the (now effectively unoccupied) level $\tilde{k}$, i.e., by a mapping
\begin{align}
    (p_{\tilde{k}-1},0) & \mapsto(\cos^{2}\!\theta\,p_{\tilde{k}-1},\sin^{2}\!\theta\,p_{\tilde{k}-1}).
    \label{smooth transition}
\end{align}
The corresponding rotation angle $\theta$ is given by
\begin{align}
    \theta  & =\,\arcsin\sqrt{\frac{\Delta\epsilon_{\mathrm{I\hspace*{-0.5pt}I}}}{p_{\tilde{k}-1}}}\,=\,\arcsin\sqrt{\frac{e^{\tilde{k}\beta\omega}\Delta\epsilon-1}{e^{\beta\omega}-1}}.
    \label{eq:min fluc theta}
\end{align}
This protocol is optimal since each (finite size) weight is shifted by either $0$ or $1$ units of energy, i.e., the shifts closest to $\Delta\epsilon$ since $0\leq\Delta\epsilon\leq1$. Explicitly, we can calculate the corresponding fluctuations by splitting the contributions for the differently shifted weights, i.e.,
\begin{align}
    (\Delta W)^{2}   &=\,(\Delta W_{<\tilde{k}-2})^{2}\,+\,(\Delta W_{\tilde{k}-1})^{2}\,+\,(\Delta W_{\geq\tilde{k}})^{2}.
\end{align}
For $n=0,\ldots,\tilde{k}-2$ we have $p_{m\rightarrow n}=p_{m}\delta_{mn}$ and $W_{m\rightarrow n}=0$ and hence $(\Delta W_{<\tilde{k}-2})^{2}=\sum_{n=0}^{\tilde{k}-2}p_{n}(\Delta\epsilon)^{2}$. For the level $\tilde{k}-1$ we have
\begin{align}
    \left(\tfrac{\Delta W_{\tilde{k}-1}}{\omega}\right)^{2} &=\,
    p_{\tilde{k}-1\rightarrow \tilde{k}-1}(\Delta\epsilon)^{2}+
    p_{\tilde{k}-1\rightarrow \tilde{k}}(1-\Delta\epsilon)^{2}\nonumber\\[1mm]
    &=\,p_{\tilde{k}-1}\bigl((\Delta\epsilon)^{2}+\sin^{2}\!\theta\,[1-2\Delta\epsilon]\bigr),
\end{align}
where we have used (\ref{smooth transition}). The remaining shifts from $\tilde{k}$ upwards give rise to $(\Delta W_{\geq\tilde{k}})^{2}=\sum_{n=\tilde{k}}^{\infty}p_{n}(1-\Delta\epsilon)^{2}=\Delta\epsilon_{\mathrm{I}}(1-\Delta\epsilon)^{2}$. When summing up these contributions, substituting $\sin^{2}\!\theta=\Delta\epsilon_{\mathrm{I\hspace*{-0.5pt}I}}/p_{\tilde{k}-1}$ from Eq.~(\ref{eq:min fluc theta}), and noting that $\Delta\epsilon=\Delta\epsilon_{\mathrm{I}}+\Delta\epsilon_{\mathrm{I\hspace*{-0.5pt}I}}$, we find
\begin{align}
    \left(\tfrac{\Delta W_{\mathrm{min}}}{\omega}\right)^{2}  &=\,\Delta\epsilon(1-\Delta\epsilon)
\end{align}
for $0\leq\Delta\epsilon\leq1$. Finally, consider the case where $\Delta\epsilon>1$. Then we perform the protocol just described, but replace $\Delta\epsilon$ with the difference $\Delta\epsilon-\lfloor\Delta\epsilon\rfloor$ to the lower integer value. The remaining energy is now an integer multiple of $\omega$ and can be gained by shifting the entire distribution upwards by $\lfloor\Delta\epsilon\rfloor$ units, whilst filling the gaps with vanishing contributions from arbitrarily high levels (as described for $\Delta\epsilon=1$ at the beginning of this section). Since the last integer shift does not add any fluctuations, we arrive at the optimal value
\begin{align}
    \left(\tfrac{\Delta W_{\mathrm{min}}}{\omega}\right)^{2}   &=\,\bigl(\Delta\epsilon-\lfloor\Delta\epsilon\rfloor\bigr)\bigl(\lceil\Delta\epsilon\rceil-\Delta\epsilon\bigr).
\end{align}
Note that this expression for the minimal fluctuations at arbitrary temperatures coincides with the expression for the minimal variance $(\Delta\sigma_{\mathrm{min}}/\omega)^{2}$ achievable at zero temperature, as given in Eq.~(\ref{eq:zero temp opt var}) and illustrated (by the lower curve) in Fig.~\ref{fig:optimal charging}, but for finite temperatures the protocol minimizing the fluctuations {does not} minimize the variance, and vice versa. {As a remark, note that in contrast to the optimal precision protocol, the protocol for minimal fluctuations does not translate directly to the finite-dimensional case.}

As for the case of the variance, {let us now turn to the case of several modes, starting with two. Here it is first important to note that a second battery can be added without increasing the fluctuations since for local unitaries one finds}
\begin{align}
    (\Delta W)^{2}[\Delta E]    &=\,(\Delta W\subtiny{0}{0}{A})^{2}[\Delta E\subtiny{0}{0}{A}]\,+\,(\Delta W\subtiny{0}{0}{B})^{2}[\Delta E\subtiny{0}{0}{B}],
\end{align}
{where $\Delta E=\Delta E\subtiny{0}{0}{A}+\Delta E\subtiny{0}{0}{B}$. Second, one may note that the protocol described above can now achieve vanishing fluctuations also for energies $\Delta E=m\omega_{A}+n\omega_{B}$ for $m,n\in\mathbb{N}_{0}$, not just for integer multiples of a single frequency. In addition, the optimization of the energy splitting between the two modes can lead to lower fluctuations as compared to only charging one of the batteries also for energy values that lie in between two choices of $m$ and $n$. All of this can be done using only local unitary charging. Correlating unitaries again only play a minor role in the sense that they may be employed in optimizing the second part of the protocol, where the missing energy $\Delta\epsilon_{\mathrm{I\hspace*{-0.5pt}I}}$ is added. The presence of multiple modes as batteries to be charged can hence be considered to be helpful.}

However, as before, the exact optimal protocols for multiple modes depend on the respective frequencies, temperatures, and on the input energy, and hence require case-by-case analyses. It thus becomes ever more clear that the operations to optimize either the variance or fluctuations are generally complicated and require extreme levels of control over the infinite-dimensional systems we consider here. It is hence of great interest to turn to practical operations such as Gaussian unitaries, and investigate their limitations for realistic battery charging.


\section{Battery Charging Using Gaussian Unitaries}\label{sec:battery charging with Gaussian unitaries}
\vspace*{-2mm}


\subsection{Preliminaries: Phase space and Gaussian states}
\vspace*{-1mm}

In the following, we want to study the restrictions imposed on the battery charging scenario when only Gaussian unitaries are used, i.e., unitary operations that map Gaussian states to Gaussian states. To examine this class of states, note that any quantum state $\rho$ in the Hilbert space $\mathcal{L}_{2}(\mathbb{R}^{N},dx)$, i.e., the space of square-integrable (with respect to the Lebesque measure $dx$) functions over $\mathbb{R}^{N}$, can be assigned a \emph{Wigner function} $\mathcal{W}(x,p)$ given by
\begin{align}
    \mathcal{W}(x,p) &=\,\frac{1}{(2\pi)^{N}}\int\!\!dy\,e^{-i\nr p\nr y}\bra{x+\tfrac{y}{2}}\rho\ket{x-\tfrac{y}{2}}\,,
    \label{eq:general Wigner function}
\end{align}
where $x,y,p\in\mathbb{R}^{N}$, with $x=(x_{1},x_{2},\ldots,x_{N})^{T}$ and $p=(p_{1},p_{2},\ldots,p_{N})$ are appropriate position and momentum coordinates and $\hat{x}=(\hat{x}_{1},\hat{x}_{2},\ldots,\hat{x}_{N})^{T}$ and $\hat{p}=(\hat{p}_{1},\hat{p}_{2},\ldots,\hat{p}_{N})^{T}$ are the corresponding position and momentum operators, respectively. The eigenstates $\ket{x}$ and $\ket{p}$ of these operators, respectively, satisfy
\begin{align}
    \hat{x}\ket{x}  &=\,x\ket{x},\\[1mm]
    \hat{p}\ket{p}  &=\,p\ket{p}.
\end{align}
It is convenient to collect $x$ and $p$ into a single phase space coordinate $\xi=(x_{1},p_{1},x_{2},p_{2},\ldots,x_{N},p_{N})^{T}\in\mathbb{R}^{2N}$, and corresponding quadrature operators $\mathds{X}_{i}$, where
\begin{align}
    \mathds{X}_{2n-1}   &=\,\hat{x}_{n}\,=\,\tfrac{1}{\sqrt{2}}\bigl(a_{n}+a_{n}^{\dagger}\bigr),\\[1mm]
    \mathds{X}_{2n} &=\,\hat{p}_{n}\,=\,\tfrac{-i}{\sqrt{2}}\bigl(a_{n}-a_{n}^{\dagger}\bigr).
\end{align}
The commutation relation $\comm{a_{m}}{a_{n}^{\dagger}}=\delta_{mn}$ then implies the canonical commutator $\comm{\hat{x}_{m}}{\hat{p}_{n}}=i\delta_{mn}$, and vice versa. For the Wigner function, the normalization of the density operator translates to the condition
\begin{align}
    \int\!\!dx\nr dp\, \mathcal{W}(x,p) &=\,\int\!\!d\xi\,\mathcal{W}(\xi)\,=\,1\,.
\end{align}
Expectation values of Hilbert space operators $\hat{G}$ can be computed from the Wigner function via
\begin{align}
    \expval{\hat{G}}_{\rho}  &=\,\tr\bigl(\hat{G}\rho\bigr)\,=\,\int\!\!dx\nr dp\,\mathcal{W}(x,p)\,g(x,p)\,,
    \label{eq:exp value via Wigner function}
\end{align}
where the Wigner transform $g(x,p)$ of the operator $\hat{G}$ is given by
\begin{align}
    g(x,p)  &=\,\int\!\!dy\,e^{i\nr p\nr y}\bra{x-\tfrac{y}{2}}\hat{G}\ket{x+\tfrac{y}{2}}.
    \label{eq:Wigner transform}
\end{align}
With these basic definitions at hand, we can now return to Gaussian states and operations.

Gaussian states are defined as those states in $\mathcal{H}$ whose Wigner function is a multivariate Gaussian, i.e., of the form
\begin{align}
    \mathcal{W}(\xi)    &=\,\tfrac{1}{\pi^{N}\sqrt{\det(\Gamma)}}\exp\bigl[-(\xi-\overline{\mathds{X}})^{T}\Gamma^{-1}(\xi-\overline{\mathds{X}})\bigr]\,,
    \label{eq:Wigner function Gaussian states}
\end{align}
for some vector $\overline{\mathds{X}}\in\mathbb{R}^{2N}$ and a real, symmetric $2N\times2N$ matrix $\Gamma$. These quantities are called the first and second statistical moments of a quantum state, where the vector of first moments is simply $\overline{\mathds{X}}=\expval{\mathds{X}}_{\rho}$ and the components of the covariance matrix are given by
\begin{align}
    \Gamma_{ij} &=\,\expval{\mathds{X}_{i}\mathds{X}_{j}+\mathds{X}_{j}\mathds{X}_{i}}\,-\,2\expval{\mathds{X}_{i}}\expval{\mathds{X}_{j}}\,.
\end{align}
Note that we have included a conventional factor of $2$ in the definition of the covariance matrix w.r.t. the actual covariances of the operators, compare, e.g., Eq.~(\ref{eq:covariance}). Via Eq.~(\ref{eq:Wigner function Gaussian states}) Gaussian states are hence fully determined by $\overline{\mathds{X}}$ and $\Gamma$.

Gaussian unitaries, which map the set of Gaussian states onto itself, are represented by affine maps $(S,\xi):\mathds{X}\mapsto S\mathds{X}+\xi$. Here $\xi\in\mathbb{R}^{2N}$ are displacements in phase space represented by the unitary Weyl operators $D(\xi)=\exp\bigl(i\mathds{X}^{T}\Omega\xi\bigr)$, which can shift the first moments, but leave the covariance matrix unchanged. The objects $S$ are real, symplectic $2N\times2N$ matrices which leave the symplectic form $\Omega$ invariant, i.e.,
\begin{align}
    S\,\Omega\,S^{T}    &=\,\Omega\,.
\end{align}
The components of $\Omega$ are given by $\Omega_{mn}=i\comm{\mathds{X}_{m}}{\mathds{X}_{n}}=\delta_{m,n-1}-\delta_{n,m+1}$. For more information on Gaussian operations and states see, e.g., Refs.~\cite{Olivares2012, Weedbrooketal2012}.


\subsection{Charging precision for single-mode Gaussian unitaries}\label{sec:charging prec Gaussian unitaries}
\vspace*{-2mm}

We now want to study the previous situation of precisely charging quantum batteries based on harmonic oscillators under the restriction to Gaussian unitaries. To this end, first note that any initial thermal state $\tau(\beta)$ is Gaussian for all temperatures (for the usual Hamiltonian $H=\sum_{n}\omega_{n}\hat{N}_{n}$ with $\hat{N}_{n}=a_{n}^{\dagger}a_{n}$). The corresponding first moments vanish, $\overline{\mathds{X}}=0$ and the covariance matrix is diagonal,
\vspace*{-2mm}
\begin{align}
    \Gamma\bigl(\tau(\beta)\bigr)   &=\,\bigoplus\limits_{n=1}^{N}\Gamma_{n}(\beta),
\end{align}
where the single-mode covariance matrices are given by $\Gamma_{n}(\beta)=\coth\bigl(\beta\omega_{n}/2\bigr)\mathds{1}_{2}$. In particular, when the temperature is zero, we have the ground state with $\Gamma_{\mathrm{vac}}=\mathds{1}$, and the corresponding Wigner function $\mathcal{W}(x,p)=\tfrac{1}{\pi}\exp[-(x^{2}+p^{2})]$.

To determine the energy of any Gaussian state for the Hamiltonian $H=\sum_{n}\omega_{n}a_{n}^{\dagger}a_{n}$ we do not need to use Eq.~(\ref{eq:exp value via Wigner function}). Inspection of the first moments $\overline{\mathds{X}}\suptiny{0.5}{-1}{(n)}=(\overline{\mathds{X}}_{2n-1},\overline{\mathds{X}}_{2n})^{T}=\expval{(\hat{x}_{n},\hat{p}_{n})^{T}}$ of each mode and local covariances simply reveals\footnote{Note that there is a typographical error in the prefactor of $\dv\overline{\mathds{X}}\suptiny{0.5}{-1}{(n)}\dv^{2}$ in Ref.~\cite[Eq.~(12)]{BrownFriisHuber2016}.} that
\vspace*{-3mm}
\begin{align}
    \hspace*{5mm}E(\rho)  &=\,\sum\limits_{n=1}^{N}
    \omega_{n}\Bigl(\frac{1}{4}\bigl[\tr(\Gamma_{n})-2\bigr]+\frac{1}{2}\dv\overline{\mathds{X}}\suptiny{0.5}{-1}{(n)}\dv^{2}\Bigr)\,.
    \label{eq:energy first and second moments}
\end{align}
However, to compute the variance {$V(\rho)$} we require also the expectation value of $H^{2}$. For our single-mode example {(for notational convenience we drop the mode label $n$ on all quantities from now on)} we have {$H=\omega\hat{N}$} with {$\hat{N}=a^{\dagger}a$}. We hence need to find the Wigner transform of {$\hat{N}^{2}$}. With some straightforward calculations which are shown in detail in Appendix~\ref{sec:wigner rep of squared number op}, one obtains the expression
\begin{align}
    N^{2}(x,p)  &=\,\tfrac{1}{4}\bigl(x^{2}+p^{2}-1\bigr)^{2}\,-\,\tfrac{1}{4}.
    \label{eq:Wigner transform of N squared main text}
\end{align}
With this, we can compute the expectation value {$\expval{\hat{N}^{2}}$} for arbitrary single-mode Gaussian states in terms of the corresponding first and second moments. Since we are dealing with a mode-local operator, we can use the single-mode version of Eq.~(\ref{eq:Wigner function Gaussian states}) to do so, i.e., using only the vector {$\overline{\mathds{X}}\in\mathbb{R}^{2}$} and the $2\times2$ covariance matrix {$\Gamma$}. After some lengthy but straightforward algebra we find that for any single-mode Gaussian state
\begin{align}
    \expval{\hat{N}^{2}}    &=\,\int\!\!dx\nr dp\,\mathcal{W}(x,p)N^{2}(x,p)\nonumber\\[1mm]
    &=\,
    \Bigl(\tfrac{1}{4}\bigl[\tr(\Gamma)-2\bigr]+\tfrac{1}{2}\dv\overline{\mathds{X}}\dv^{2}\Bigr)^{2}\nonumber\\[1mm]
    &\ +\,\tfrac{1}{2}\overline{\mathds{X}}\phantom{}^{T}
    \Gamma\,
    \overline{\mathds{X}}
    \,+\,\tfrac{1}{8}\bigl[\tr(\Gamma^{2})-2\bigr]
    \,.
    \label{eq:exp value N squared first and second moments}
\end{align}
Since the first term on the right-hand side of Eq.~(\ref{eq:exp value N squared first and second moments}) is just the squared expectation value of {$\hat{N}$} for a Gaussian state [compare with Eq.~(\ref{eq:energy first and second moments})], we immediately obtain the variance
\begin{align}
    (\Delta \hat{N})^{2}    &=\,\tfrac{1}{2}\overline{\mathds{X}}\phantom{}^{T}
    \Gamma\,\overline{\mathds{X}}
    \,+\,\tfrac{1}{8}\bigl[\tr(\Gamma^{2})-2\bigr].
    \label{eq:variance of N first and second moments}
\end{align}

With this knowledge at hand, we can now return to our problem of raising the energy of a single-mode battery using Gaussian unitaries. For single-mode batteries that are initially at a finite temperature, the initial energy and corresponding variance can be calculated from Eqs.~(\ref{eq:energy first and second moments}) and~(\ref{eq:variance of N first and second moments}) by noting that the corresponding first moments vanish, {$\overline{\mathds{X}}=0$} and the covariance matrix is {$\Gamma(\beta)=\coth\bigl(\beta\omega/2\bigr)\mathds{1}_{2}$}. With this we have
\begin{align}
    E\bigl(\tau(\beta)\bigr)    &=\,\tfrac{\omega}{2}\bigl[\coth\bigl(\tfrac{\beta\omega}{2}\bigr)-1\bigr]\,,
    \label{eq:single mode thermal state energy}\\[1mm]
    V(\tau)\,=\,\bigl(\Delta H(\tau)\bigr)^{2}  &=\,\tfrac{\omega^{2}}{4}\bigl[\coth^{2}\!\bigl(\tfrac{\beta\omega}{2}\bigr)-1\bigr]\,.
    \label{eq:single mode thermal state variance}
\end{align}

\begin{figure}[ht!]
\label{fig:single mode Gaussian unitaries ground state}
\begin{center}
\includegraphics[width=0.48\textwidth,trim={0cm 0mm 0cm 0mm}]{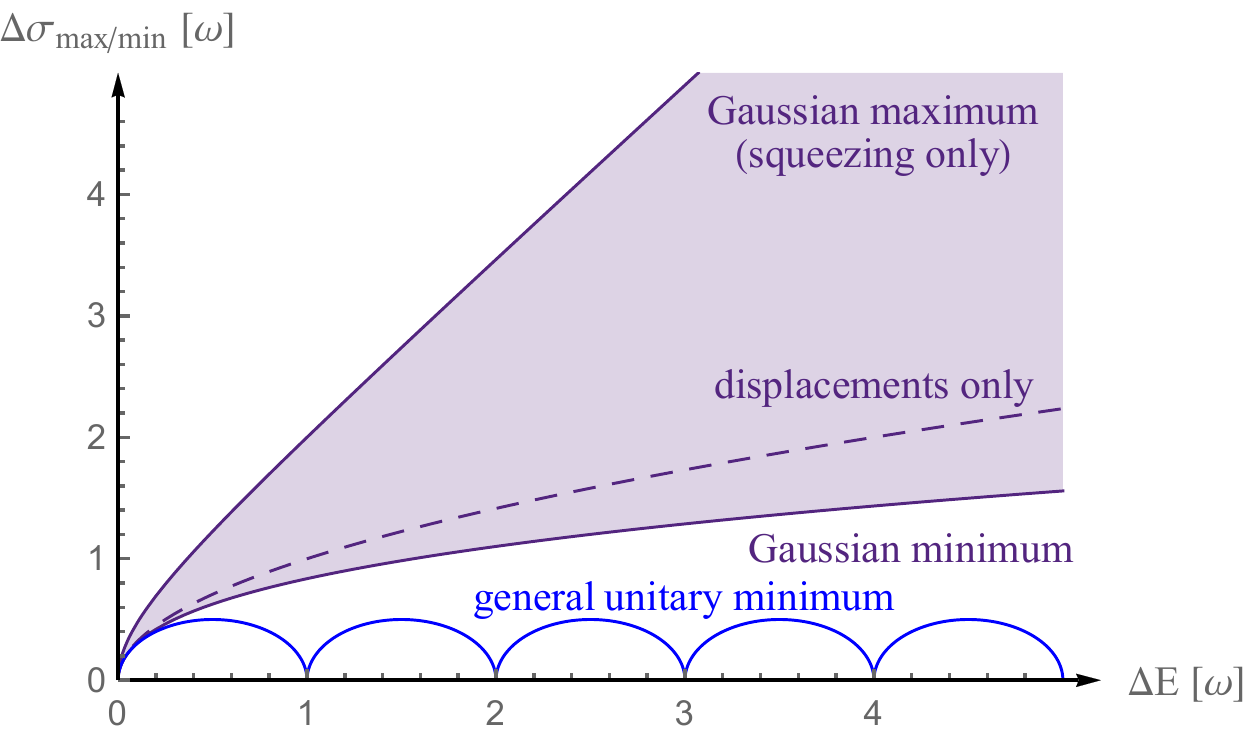}
\vspace*{-6mm}
\end{center}
\caption{
\textbf{Ground state battery charging:} The maximal and minimal variances of the energy that are possible using Gaussian unitaries for a battery that has been charged by $\Delta E$ and starting in its ground state are shown (in units of $\omega$, with $\hbar=1$). For reference, the performance of pure displacements and the lower bound for arbitrary unitaries are also shown.}\vspace*{-2mm}
\end{figure}

We can then apply Gaussian unitaries to these states. For instance, we may consider single-mode displacements to raise the energy of initial thermal states. For vanishing temperature, the action of the corresponding Weyl displacement operators $D(\xi)$ on the vacuum creates coherent states $D(\xi)\ket{0}=\ket{\alpha}=e^{-|\alpha|^{2}/2}\sum_{j}\tfrac{\alpha^{j}}{\sqrt{j!}}\ket{j}$, where $\xi=\sqrt{2}(\operatorname{Re}(\alpha),\operatorname{Im}(\alpha))^{T}\in\mathbb{R}^{2}$ and $\alpha\in\mathbb{C}$. Since displacements do not alter the covariance matrix, the latter remains that of a single-mode thermal state, while the first moments are transformed to $\overline{\mathds{X}}_{i}=\xi_{i}$. We hence have
\begin{subequations}
\begin{align}
    \tfrac{\Delta E}{\omega}   &=\,\tfrac{1}{2}\dv\overline{\mathds{X}}\dv^{2}\,=\,|\alpha|^{2}\,,\\[1mm]
    (\Delta\hat{N})^{2} &=\,
    \tfrac{1}{2}\coth(\tfrac{\beta\omega}{2})\dv\overline{\mathds{X}}\dv^{2}+
    \tfrac{V(\tau)}{\omega^{2}}
    .
\end{align}
\end{subequations}
For displaced thermal states we consequently find
\begin{align}
    \tfrac{\Delta \sigma}{\omega}   &=\,\sqrt{
    \coth(\tfrac{\beta\omega}{2})\tfrac{\Delta E}{\omega}+
    \tfrac{V(\tau)}{\omega^{2}}
    }
    \,-\,\sqrt{\tfrac{V(\tau)}{\omega^{2}}
    },
    \label{eq:delta sigma displacements}
\end{align}
i.e., an asymptotic increase of the energy standard deviation with the square-root of the energy increase. As we shall see, pure displacements are neither optimal (minimal $\Delta\sigma$ for given $\Delta E$), nor the worst possible Gaussian operations for battery charging, but nonetheless, make for an interesting comparison. This is illustrated in Fig.~\ref{fig:single mode Gaussian unitaries ground state}, where we have also included results for the optimal and worst operations, which we shall derive next.


\subsection{Optimal and worst-case Gaussian charging precision}\label{sec:Optimal and worst-case Gaussian charging precision}

Let us now investigate these best-case and worst-case Gaussian operations. The action of an arbitrary local Gaussian unitary $U\subtiny{-1}{-1}{G}$ results in some (generally nonzero) first moments {$\xi=\overline{\mathds{X}}(U\subtiny{-1}{-1}{G}\tau U\subtiny{-1}{0}{G}^{\dagger})$}, while {$\Gamma(\beta)$} is mapped to the covariance matrix {$\tilde{\Gamma}$} of an arbitrary single-mode Gaussian state with the same mixedness via a local symplectic operation $S_{\mathrm{loc}}$, i.e., $\tilde{\Gamma}=S_{\mathrm{loc}}\Gamma S_{\mathrm{loc}}^{T}$. Any single-mode symplectic operation can be decomposed~\cite{Braunstein2005, Weedbrooketal2012} into (phase) rotations $R$ and single-mode squeezing transformations $S(r)$ as
\begin{align}
    S_{\mathrm{loc}}    &=\,R(\theta)\,S(r)\,R(\phi)\,,
\end{align}
where $\theta, \phi$ are real rotation angles, $r\in\mathbb{R}$ is the squeezing parameter, and
\begin{align}
        R(\theta)=\begin{pmatrix}   \cos\theta  &   \sin\theta  \\  -\sin\theta &   \cos\theta  \end{pmatrix}\,,\ \
        S(r)    \,=\,\begin{pmatrix}   e^{-r}  &   0  \\  0 &   e^{r}  \end{pmatrix}\,.
\end{align}
With this, the transformed covariance matrix can be written as
\begin{align}
    \tilde{\Gamma}  &=\,\coth\bigl(\tfrac{\beta\omega}{2}\bigr)\,R(\theta)\,S(2r)\,R(\theta)^{T}\,,
\end{align}
and the corresponding average energy of the state {$\rho=U\subtiny{-1}{-1}{G}\tau U\subtiny{-1}{0}{G}^{\dagger}$} evaluates to
\begin{align}
    E(\rho)    &=\,
    \tfrac{\omega}{2}\bigl[\coth\bigl(\tfrac{\beta\omega}{2}\bigr)\,\cosh(2r)-1+\dv\xi\dv^{2}\bigr]\,.
\end{align}

Combining this with Eq.~(\ref{eq:single mode thermal state energy}), we then have the energy input
\begin{align}
    \Delta E    &=\,E(\rho)\,-\,E(\tau) \nonumber\\[1mm]
    &=\,\tfrac{\omega}{2}\bigl[\coth\bigl(\tfrac{\beta\omega}{2}\bigr)\,\bigl(\cosh(2r)-1\bigr)+\dv\xi\dv^{2}\bigr]\,.
    \label{eq:Delta E for single mode Gaussian states}
\end{align}
For the variance of the energy of the final state we first inspect the term {$\xi^{T}\tilde{\Gamma}\xi$} and note that for our intents we can absorb the rotation $R(\theta)$ into the choice of the first moments, since $\dv\xi\dv^{2}=\dv R^{T}(\theta)\xi\dv^{2}$. We hence find
\begin{align}
    V(\rho)=\bigl(\Delta H(\rho)\bigr)^{2}  &=\,
    \tfrac{\omega^{2}}{4}\bigl[\coth^{2}\!\bigl(\tfrac{\beta\omega}{2}\bigr)\cosh(4r)-1\nonumber\\[1mm]
    &\ +\,2\coth\bigl(\tfrac{\beta\omega}{2}\bigr)\,\xi^{T}S(2r)\xi\,\bigr]\,.
    \label{eq:Vrho for single mode Gaussian states}
\end{align}

{We then proceed in the following way. First, we note that $\Delta E$ is a function of $r$ and $\dv\xi\dv^{2}$, whereas $V(\rho)$ depends on $\xi$ only via the term $\xi^{T}S(2r)\xi=\xi_{1}^{2}e^{-2r}+\xi_{2}^{2}e^{2r}$. Since $e^{-2r}\leq e^{2r}$ for $r\geq0$, the maximal and minimal values of $V(\rho)$ for fixed $\Delta E$ and $T=1/\beta$ must be attained for combinations of $r\geq0$ and $\xi$ with $\xi_{1}=0$ and $\xi_{2}=0$, respectively. Conversely, this means that the remaining quantities $\xi_{2}^{\,2}$ and $\xi_{1}^{\,2}$ in Eq.}~(\ref{eq:Vrho for single mode Gaussian states}) {can be identified with $\dv\xi\dv^{2}$, i.e., from  Eq.}~(\ref{eq:Delta E for single mode Gaussian states}) {we have}
\begin{align}
    \tfrac{2\Delta E}{\omega}-\coth\bigl(\tfrac{\beta\omega}{2}\bigr)\!\bigl(\cosh(2r)-1\bigr) &=
    \begin{cases}
        \xi_{2}^{\,2}\ \text{if $V$ maximal},\\
        \xi_{1}^{\,2}\ \text{if $V$ minimal}.
    \end{cases}
\end{align}
{We can thus define two functions $V_{+}(r)$ and $V_{-}(r)$, given by}
\begin{align}
    \tfrac{V_{\pm}(r)}{\omega^{2}} &=\,\tfrac{1}{4}\bigl[\coth\bigl(\tfrac{\beta\omega}{2}\bigr)^{2}\cosh(4r)-1\bigr]
    \label{eq:VmaxVmin}\\
    &\ +e^{\pm2r}\bigl[\tfrac{\Delta E}{\omega}-\tfrac{1}{2}\coth\bigl(\tfrac{\beta\omega}{2}\bigr)\!\bigl(\cosh(2r)-1\bigr)\bigr],
    \nonumber
\end{align}
{which correspond to the respective restrictions of the final state variance $V(\rho)$. Moreover, when the initial temperature and input energy are fixed these functions depend only on $r$, allowing one to straightforwardly determine the maxima $\max_{r} V_{+}(r)=\max_{r}V$ and minima $\min_{r} V_{-}(r)=\min_{r}V$, respectively. As we show in detail in Appendix}~\ref{sec:Extremal Gaussian Precision and Fluctuations}{, for every fixed $\Delta E\geq0$ and $T\geq0$, there exist unique values $r_{\pm}\geq0$, such that $\max_{r} V_{+}(r) =V_{+}({\rplus} )=\max_{r}V=V({\rplus} )$ and $\min_{r} V_{-}(r) =V_{-}({\rminus} )=\min_{r}V=V({\rminus} )$.}



\subsubsection{Worst-case Gaussian charging precision}\label{sec:worst-case Gaussian charging precision}

{In particular, we find (see Appendix}~\ref{sec:max variance} {for details) that for any temperature and energy input, the maximal values of $\Delta\sigma$} are obtained for $\xi=0$, that is, when all energy is transferred to the battery via single-mode squeezing. In this case both $\Delta E$ and $\Delta\sigma$ are functions of $r$ only, and we can {use Eq.}~(\ref{eq:Delta E for single mode Gaussian states}){to relate the two quantities directly. In other words, we find}
\begin{align}
    {\rplus}    &=\,\tfrac{1}{2}\operatorname{arcosh}\left(2\tfrac{\Delta E}{\omega}/\coth(\beta\omega/2)+1\right)\,.\,
    \label{eq:squeezing parameter rplus}
\end{align}
{along with the maximal variance increase}
\begin{align}
    \tfrac{\Delta \sigma_{\mathrm{max}}}{\omega}    &=\sqrt{2\tfrac{\Delta E}{\omega}\bigl(\tfrac{\Delta E}{\omega}+\coth(\tfrac{\beta\omega}{2})\bigr)
    +\!\tfrac{V(\tau)}{\omega^{2}}
    }
    -\sqrt{\tfrac{V(\tau)}{\omega^{2}}
    }.
    \label{eq:max Gaussian standard dev increase}
\end{align}
As we see, {in this case the energy standard deviation increases linearly with the energy input in the asymptotic regime (as $\Delta E\rightarrow\infty$)}.

\subsubsection{Optimal Gaussian charging precision}\label{sec:optimal Gaussian charging precision}

{While the worst-case Gaussian unitary transformation has thus been identified as pure single-mode squeezing, the Gaussian unitary transformation that minimizes the variance of the energy can be identified as a combination of squeezing and displacement that depends on the input energy and temperature, see Appendix}~\ref{sec:min variance}{. That is, in our conventions, the optimal performance is achieved for $\xi_{2}=0$ and generally nonzero values of $\xi_{1}$ and $r={\rminus} $, where the latter is determined by the condition $\left.\partial V_{-}/\partial r\right|_{r={\rminus} } = 0$, which implies}
\begin{align}
    \tfrac{\Delta E}{\omega}    &=\,\tfrac{1}{2}\coth(\tfrac{\beta\omega}{2})\bigl(e^{2{\rminus} }\cosh(4{\rminus} )-1\bigr)\,.
    \label{eq:r implicit}
\end{align}
{Inserting Eq.}~(\ref{eq:r implicit}) {into $V_{-}$ from Eq.}~(\ref{eq:VmaxVmin}) {then permits us to write}
\begin{align}
    \tfrac{V_{-}(\rminus)}{\omega}
    &=
    \tfrac{1}{2}\!\coth^{2}\!(\!\tfrac{\beta\omega}{2}\!)\bigl[e^{2{\rminus} }\!\sinh(2{\rminus} )\!+\!\tfrac{1}{2}\bigl(\cosh(4{\rminus} )\!-\!1\bigr)\bigr].
    \label{eq:delta sigma as function of r}
\end{align}
{Although a closed expression for ${\rminus} (\Delta E, \beta)$ cannot be given, we show in Appendix}~\ref{sec:min variance} {that ${\rminus} (\Delta E, \beta)\geq0$ exists and is unique and can thus easily be determined numerically for any given $\Delta E$ and $T=1/\beta$ via the implicit formula in Eq.}~(\ref{eq:r implicit}). {The value ${\rminus} $ obtained in this way can then be inserted}
\begin{figure}[ht!]
\label{fig:comparison of charging protocols}
\begin{center}
\includegraphics[width=0.48\textwidth,trim={0cm 0mm 0cm 0mm}]{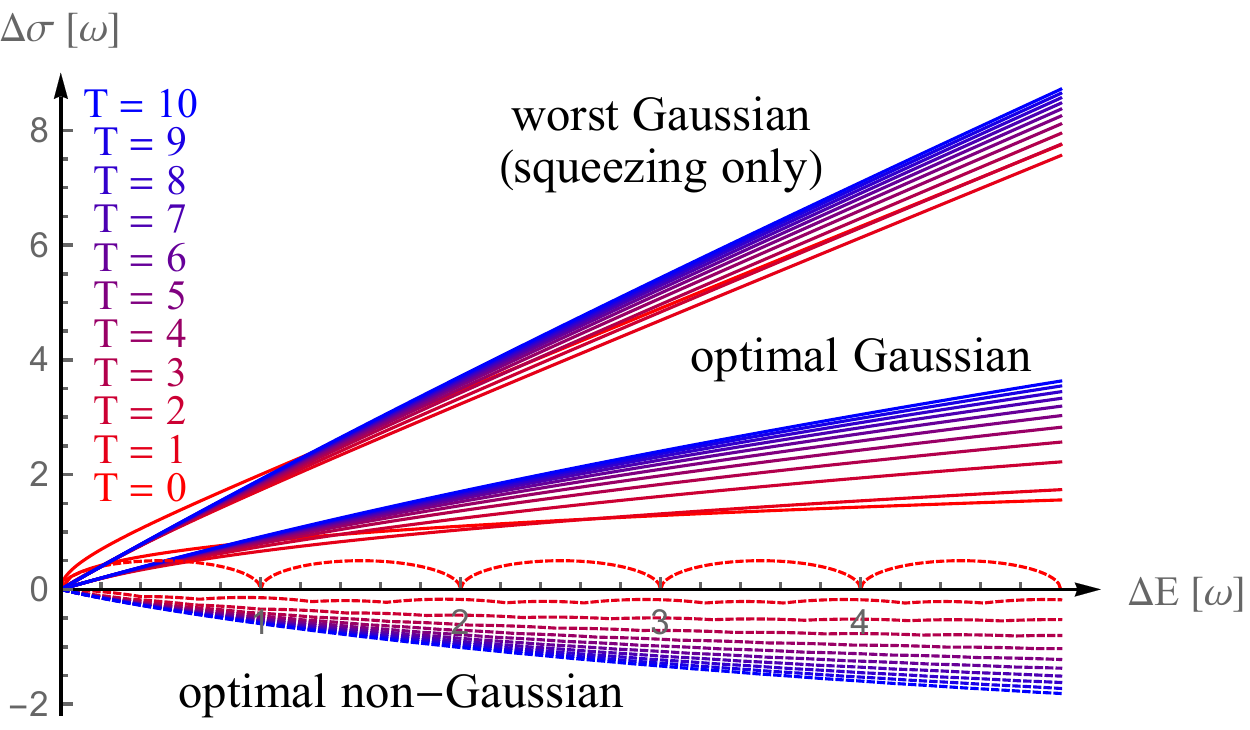}
\vspace*{-2mm}
\end{center}
\caption{
\textbf{Gaussian vs. optimal precision charging:} {The standard deviation change $\Delta\sigma$ (in units of {$\omega$}) is shown as a function of the input energy $\Delta E$ (in units of {$\omega$}) for the worst-case (squeezing only, upper group of curves) and optimal (middle group of curves) single-mode Gaussian unitaries, as compared with the corresponding optimal non-Gaussian values (lower group of curves) obtained via the optimal protocol discussed in Section}~\ref{sec:Fundamental limitations thermal states single mode}{. Each group of curves corresponds to temperatures $T=0$ to $10$ in steps of $1$ (in units of $\omega$).}
}\vspace*{-2mm}
\end{figure}
\noindent
{into Eq.}~(\ref{eq:delta sigma as function of r}) {to arrive at the minimal variance. Results for the bounds on $\Delta\sigma$ for a initial thermal states are shown in Fig.}~\ref{fig:comparison of charging protocols} {for a range of temperatures and input energies.}

Here it is interesting to note that, while the upper bound is achieved for pure squeezing transformations, the lower bound is a combination of squeezing and displacements. In the optimal case, the energies $\Delta E_{\mathrm{sq}}$ and $\Delta E_{\mathrm{d}}$ invested into squeezing and displacement, respectively, are expressed via the optimal squeezing parameter {${\rminus} $} as
\begin{align}
    \tfrac{\Delta E_{\mathrm{d}}}{\omega}  &=\,\tfrac{1}{2}\xi_{1}^{2}\,=\,\tfrac{1}{2}\coth(\tfrac{\beta\omega}{2})e^{4{\rminus} }\sinh(2{\rminus} )\,,\\[1mm]
    \tfrac{\Delta E_{\mathrm{sq}}}{\omega}  &=\,\tfrac{\Delta E-\Delta E_{\mathrm{d}}}{\omega}\,=\,\tfrac{1}{2}\coth(\tfrac{\beta\omega}{2})\bigl(\cosh(2{\rminus} )-1\bigr)\,.
\end{align}
In the limit of large energy supplies, $\Delta E\rightarrow\infty$ (at fixed temperature this implies {${\rminus} \rightarrow\infty$}), we hence have
    $\tfrac{\Delta E_{\mathrm{d}}}{\Delta E_{\mathrm{sq}}}\rightarrow\,e^{4{\rminus} }$.
That is, the energy invested into squeezing grows much less strongly than that invested into displacements.

Moreover, note that while pure displacement asymptotically (i.e., for $\Delta E\rightarrow\infty$) leads to a linear scaling of the final variance with $\Delta E$, {that is, $V(\rho)/\Delta E\rightarrow\omega\coth(\tfrac{\beta\omega}{2})$ as $\Delta E\rightarrow\infty$, see Eq.~}(\ref{eq:delta sigma displacements}){, the optimal local strategy provides a more favourable scaling behaviour even though most of the energy is invested into displacement. More specifically, considering the relative variance $V_{-}/\Delta E$ by combining Eqs.}~(\ref{eq:r implicit}) {and}~(\ref{eq:delta sigma as function of r}){, and taking the limit $\Delta E\rightarrow\infty$ (corresponding to $\rminus\rightarrow\infty$), one finds that $V_{-}/\Delta E\rightarrow0$, i.e., the optimal variance scales sub-linearly with the input energy.}



\subsection{Charging precision for multi-mode Gaussian unitaries}
\vspace*{-2mm}

Let us now finally turn to the case of bounding the charging precision of a multi-mode battery under the restriction to Gaussian unitaries. As we have discussed in Section~\ref{sec:Fundamental limitations multi mode}, in a general optimal protocol for the charging precision, correlations between the individual battery systems can be helpful in principle. However, this appears to be the case only if one can selectively rotate between specific energetically desirable levels. For Gaussian unitaries, such specialized operations with, in a manner of speaking, surgical precision are out of the question. In particular, one may view any multi-mode Gaussian operation as a combination of local operations and beam splitting~\cite{Braunstein2005}. The latter may shift the average excitation numbers between the modes but may do little more. Another aspect of introducing correlations during the charging process is that any energy stored in this way also has to be extracted globally from the joint system if optimality is to be preserved. In other words, introducing correlations raises the effective local temperatures (and hence the local entropies), reducing the local free energy. In the spirit of restricting to practical operations, we shall hence consider only local Gaussian operations from now on.

Nonetheless, it may sometimes be useful to split the energy supply between different modes in specific ways, depending on the initial temperature and energy supply. To understand why it is useful, consider the (non-optimal) case of charging two modes labelled $A$ and $B$ {(with frequencies $\omega_{A}$ and $\omega_{B}$, respectively)} via pure displacements. For such local operations, no correlations are introduced. If the energy is split in such a way that for some real $p\in [0,1]$ the energy $p\Delta E$ is stored in the mode $A$ and $(1-p)\Delta E$ in the mode $B$, inspection of Eq.~(\ref{eq:delta sigma displacements}) reveals that the variance of the final state $\rho\subtiny{0}{0}{AB}$ behaves as
\begin{align}
    V(\rho\subtiny{0}{0}{AB})   &=\bigl(p\nu\subtiny{0}{0}{A}\omega\subtiny{0}{0}{A}\!+\!(1\!-\!p)\nu\subtiny{0}{0}{B}\omega\subtiny{0}{0}{B}\bigr)\Delta E\!+\!V(\tau\subtiny{0}{0}{A})\!+\!V(\tau\subtiny{0}{0}{B})
\end{align}
with $\nu_{i}=\coth\bigl(\tfrac{\beta\omega_{i}}{2}\bigr)$ for $i=A,B$. When the two modes have the same frequencies, $\omega\subtiny{0}{0}{A}=\omega\subtiny{0}{0}{B}$, the variance becomes independent of $p$, i.e., {$V(\rho\subtiny{0}{0}{AB})=\nu\subtiny{0}{0}{A}\omega\subtiny{0}{0}{A}\Delta E+2V(\tau\subtiny{0}{0}{A}) =
\nu\subtiny{0}{0}{B}\omega\subtiny{0}{0}{B}\Delta E+2V(\tau\subtiny{0}{0}{B})$}, and it does not matter how the energy is split. Otherwise, that is, when $\omega\subtiny{0}{0}{A}\neq\omega\subtiny{0}{0}{B}$, it becomes beneficial to store all the energy in the lower frequency mode. Now, recall that this is the case for pure displacements, which are not optimal. The optimal strategy, in contrast, provides an increase of the variance that is sub-linear with the input energy. In this case it matters how the energy is split for all frequency combinations. E.g., for $\omega\subtiny{0}{0}{A}=\omega\subtiny{0}{0}{B}$ it becomes optimal to evenly divide the energy between the two batteries. In general, the optimal energy per battery is determined by the number and temperature of the batteries and their respective frequencies.

The worst case local scenario is obtained when the energy is used only for single-mode squeezing, where the splitting of the energy between the modes again depends on the specific situation. For instance, when the frequencies of both modes are the same, investing the same energy in both batteries via squeezing will lead to the largest variance. This is because the local variances increase stronger than linearly with the input energy for single-mode squeezing.


\vspace*{-1mm}
\subsection{Charging fluctuations for single-mode Gaussian unitaries}\label{sec:Charging fluctuations for single-mode Gaussian unitaries}
\vspace*{-2mm}

\begin{figure}[ht!]
\label{fig:comparison of fluctuation protocols}
\begin{center}
\includegraphics[width=0.48\textwidth,trim={0cm 0mm 0cm 0mm}]{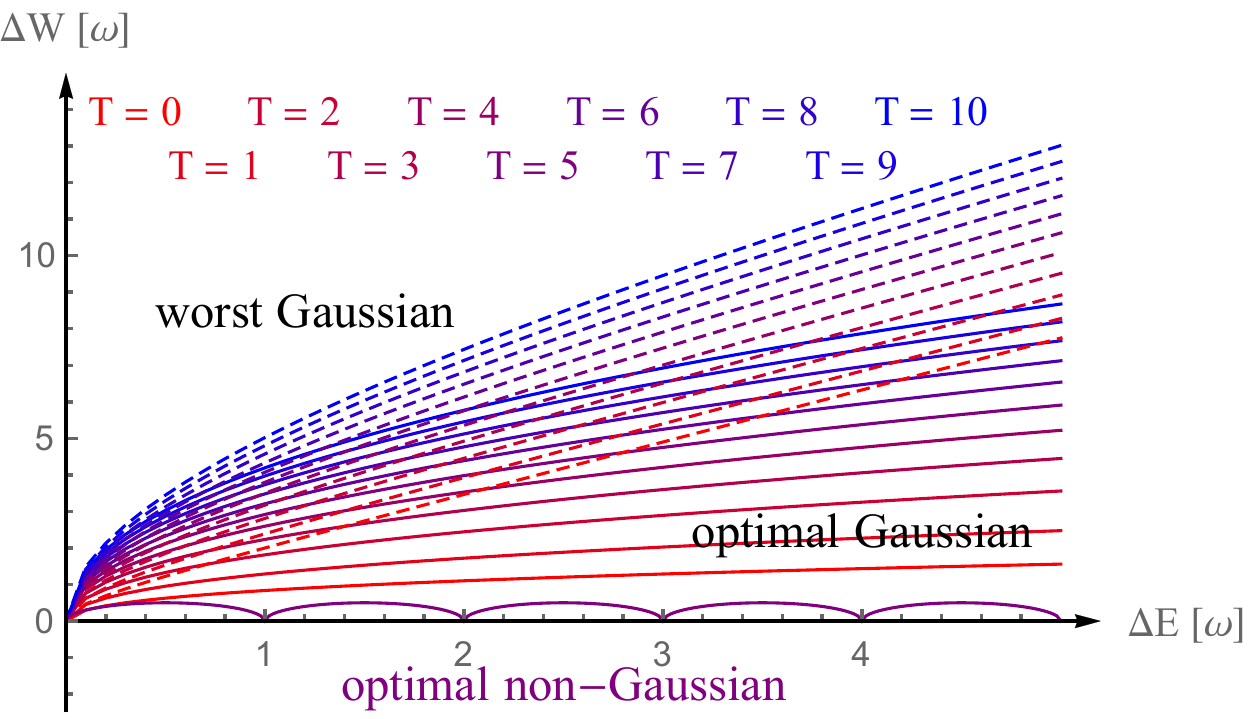}
\end{center}
\vspace*{-2mm}
\caption{
\textbf{Gaussian vs. optimal fluctuation charging:} The minimal (solid) and maximal (dashed) fluctuations $\Delta W$ (in units of {$\omega$}) achievable with Gaussian unitaries are shown for charging quantum batteries initially at temperatures $T=0$ (red) to $T=10$ (blue) (in steps of $1$) for given energy input $\Delta E$ (in units of {$\omega$}). The periodic purple curve at the bottom indicates the minimal fluctuations that are in principle achievable, as described in Section~\ref{sec:Fundamental limitations fluctuations thermal states single mode}.}
\end{figure}

At last, let us turn to the question of bounding the possible fluctuations that may appear in Gaussian battery charging. From Eq.~(\ref{eq:squared work fluctuation}) we already know how to express the energies and variances of $\rho$ and $\tau$ in terms of the temperature and final first and second moments. However, we still need to calculate $\tr(\tilde{H}H\tau)$ for arbitrary Gaussian unitaries, where we restrict to local operations, as before. For a single mode {with frequency $\omega$} we may write the term in question as
\begin{align}
    \tfrac{1}{\omega^{2}}\tr(\tilde{H}H\tau) & 
    =\!\sum\limits_{n}p_{n}\,n\bra{n}U^{\dagger}a^{\dagger}a\,U\ket{n}\,.
    \label{eq:fluctuations matrix element}
\end{align}
In principle, an arbitrary single-mode Gaussian unitary $U\subtiny{-1}{-1}{G}$ may be decomposed into a combination of single-mode squeezing operations, local rotations, and displacements, none of which commute with each other. In spite of this, for any unitary and any initial state $\rho_{o}$ we can find a single-mode Gaussian unitary $\tilde{U}\subtiny{-1}{-1}{G}=D(-\xi)U\subtiny{-1}{-1}{G}$ such that the first moments of $\tilde{U}\subtiny{-1}{-1}{G}\rho_{o}\tilde{U}\subtiny{-1}{0}{G}^{\dagger}$ vanish. Conversely, this means that for the purpose of calculating $\bra{n}U^{\dagger}a^{\dagger}a\,U\ket{n}$ we may assume that $U\subtiny{-1}{-1}{G}$ may be written as $U\subtiny{-1}{-1}{G}=D(\xi)\tilde{U}\subtiny{-1}{-1}{G}$, where $D(\xi)$ is a pure displacement and $\tilde{U}\subtiny{-1}{-1}{G}$ leaves the origin of the phase space invariant. Consequently, we can use the Bloch-Messiah decomposition~\cite{Braunstein2005} to write $\tilde{U}\subtiny{-1}{-1}{G}$ as a combination of single-mode squeezing $U_{\mathrm{S}}(r)$ and local rotations $R(\theta)$, i.e.,
\begin{align}
    \tilde{U}\subtiny{-1}{-1}{G}  &=\,R(\theta)U_{\mathrm{S}}(r)R(\phi).
\end{align}
Inserting into Eq.~(\ref{eq:fluctuations matrix element}) we then find that the rotations either act on the rotationally invariant Fock states (the phases cancel), or can be absorbed into the direction of the displacement (using the same symbol $\xi$ in a slight abuse of notation). We hence find
\begin{align}
    \bra{n}\nl U\subtiny{-1}{0}{G}^{\dagger}a^{\dagger}a\,U\subtiny{-1}{-1}{G}\nl \ket{n}    &\!=\!
    \bra{n}\nl U_{\mathrm{S}}(r)^{\dagger}D(\xi)^{\dagger}a^{\dagger}a\,D(\xi)U_{\mathrm{S}}(r)\nl\ket{n}.\nonumber
\end{align}
We then use the simple relations
\begin{subequations}
\begin{align}
    U_{\mathrm{S}}(r)^{\dagger}a\,U_{\mathrm{S}}(r)   &=\,\cosh(r)\,a\,+\,\sinh(r)\,a^{\dagger},\\
    D(\xi)^{\dagger}a\,D(\xi)  &=\,a\,+\,\tfrac{\xi}{\sqrt{2}}\,
\end{align}
\end{subequations}
to obtain the desired matrix element
\begin{align}
    \bra{n}U\subtiny{-1}{0}{G}^{\dagger}a^{\dagger}a\,U\subtiny{-1}{-1}{G}\ket{n}  &=\,
    n\,\cosh(2r)\,+\,\sinh^{2\!}(r)\,+\,\tfrac{1}{2}\dv\xi\dv^{2}.
\end{align}
We can then reinsert this result into Eq.~(\ref{eq:fluctuations matrix element}) and evaluate the sum over $n$. Further inserting into the squared work fluctuations of Eq.~(\ref{eq:squared work fluctuation}), and combining this with the expressions for the variances and average energies from Eqs.~(\ref{eq:single mode thermal state energy}),~(\ref{eq:single mode thermal state variance}), and~(\ref{eq:Delta E for single mode Gaussian states}) we obtain
\begin{align}
    \left(\tfrac{\Delta W}{\omega}\right)^{2}     &=\tfrac{V(\rho)}{\omega^{2}}\!+\!\tfrac{V(\tau)}{\omega^{2}}\!-\!2\tfrac{E(\tau)}{\omega}\bigl(1\!+\!\tfrac{E(\tau)}{\omega}\cosh(2r)\bigr),
    \label{eq:single mode Gaussian fluctuations}
\end{align}
where $V(\rho)$ is given by Eq.~(\ref{eq:Vrho for single mode Gaussian states}). {Here, we note that, apart from $V(\rho)$, no dependency on the displacement $\xi$ appears in Eq.}~(\ref{eq:single mode Gaussian fluctuations}){. Consequently, we may argue as in Section}~\ref{sec:charging prec Gaussian unitaries}, {that for any given value of $r$, the maximal and minimal function values (here of $(\Delta W/\omega)^{2}$) for fixed initial temperature and fixed $\Delta E$ are attained for combinations of (single-mode) squeezing $r\geq0$ and displacements $\xi$ with $\xi_{1}=0$ and $\xi_{2}=0$, respectively. In other words, we are interested in determining the maximum of $\Delta W_{+}(r)$ and the minimum of $\Delta W_{-}(r)$, where}
\begin{align}
    \left(\tfrac{\Delta W_{\pm}(r)}{\omega}\right)^{2}
    &=\tfrac{V_{\pm}(r)}{\omega^{2}}\!+\!\tfrac{V(\tau)}{\omega^{2}}\!-\!2\tfrac{E(\tau)}{\omega}\bigl(1\!+\!\tfrac{E(\tau)}{\omega}\cosh(2r)\bigr),
    \label{eq:DW max min}
\end{align}
{and $V_{\pm}(r)$ is given by Eq.}~(\ref{eq:VmaxVmin}){. As we show in detail in Appendices}~\ref{sec:max fluctuations} {and}~\ref{sec:min fluctuations}{, both extremal values exist and are unique for any given initial temperature $T=1/\beta$ and input energy $\Delta E\geq0$. Once again, the corresponding extremal squeezing parameters $\rtpm$ (which are in general different from the optimal squeezing parameters $\rpm$ for the charging precision) are only given implicitly, i.e., by the conditions $\left.\partial \Delta W_{\pm}/\partial r\right|_{r=\rtpm}= 0$, which can be expressed as}
\begin{align}
    \tfrac{\Delta E}{\omega}    &=\,\tfrac{1}{2}\Bigl[ \coth(\beta\omega/2) \bigl(e^{\mp2{\rtpm} } \cosh(4{\rtpm} )-1\bigr)
    \label{eq:delta E fluctuations extremal}\\
    &\ \ \ \ \ \ \  \pm\tfrac{(\coth(\beta\omega/2)-1)^{2}}{\coth(\beta\omega/2)}\, e^{\mp2\rtpm} \sinh(2\rtpm)\Bigr],\nonumber
\end{align}
{respectively. Nonetheless, $\rtpm$ and hence the exact optimal and worst single-mode Gaussian fluctuations can easily be obtained numerically, which we have done for some sample values shown in Fig.}~\ref{fig:comparison of fluctuation protocols}. Interestingly, the additional terms appearing in Eq.~(\ref{eq:single mode Gaussian fluctuations}) besides the variances lead not only to a different optimum in terms of the relative strengths of squeezing and displacements, but also mean that the worst case is now also attained for nonzero displacements. {In particular, we find that the energy $\Delta E_{\mathrm{d}}^{\pm}$ invested into displacement in the extremal cases is given by}
\begin{align}
    \tfrac{\Delta E_{\mathrm{d}}^{\pm}}{\omega}  &=\tfrac{1-e^{\mp2{\rtpm}}}{2}
    \bigl[\tfrac{(\coth(\beta\omega/2)-1)^{2}}{\coth(\beta\omega/2)} - \coth(\tfrac{\beta\omega}{2})e^{\mp2{\rtpm}}\bigr],
\end{align}
\noindent 
while the energy invested into squeezing is $\Delta E_{\mathrm{sq}}^{\pm}=\Delta E-\Delta E_{\mathrm{d}}^{\pm}$.  In the limit of large energy supplies, i.e., $\Delta E\rightarrow\infty$ (corresponding to $\rtpm\rightarrow\infty$ at fixed temperature), we have
\begin{align}
    \lim_{\rtplus\rightarrow\infty}\tfrac{\Delta E_{\mathrm{d}}^{+}}{\omega}  &=\frac{(\coth(\beta\omega/2)-1)^{2}}{4\coth(\beta\omega/2)}=\mathrm{const.},
\end{align}
while $\Delta E_{\mathrm{sq}}^{+}\rightarrow\infty$. One thus finds that the worst-case Gaussian strategy invests almost all energy into squeezing asymptotically. Conversely, for the minimal fluctuations we have $\Delta E_{\mathrm{d}}^{-}/\Delta E_{\mathrm{sq}}^{-}\rightarrow e^{-4\rtminus}$ as $\rtminus\rightarrow\infty$. In the limit of large input energies it is thus optimal to invest almost all energy into displacement to minimize the fluctuations.

The crucial feature to note is that the optimal Gaussian strategy results in a sub-linear increase of the fluctuations with the input energy. {Similarly as} before for the Gaussian strategy optimizing the charging precision, {we can consider the limit $\Delta E\rightarrow\infty$ of the relative fluctuations $\Delta W_{-}^{2}/\Delta E$. For the optimal strategy, Eq.}~(\ref{eq:delta E fluctuations extremal}) {tells us that this corresponds to the limit $\rtminus\rightarrow\infty$, for which $\Delta W_{-}^{2}/\Delta E(\rtminus)\rightarrow0$, since $\Delta W_{-}^{2}$ and $\Delta E$ grow as $e^{4\rtpm}$ and $e^{6\rtpm}$, respectively, in this limit.}


Finally, let us briefly comment on the Gaussian multi-mode scenario. Much like before for the charging precision, using Gaussian operations that generate correlations seems to be practically irrelevant since any energy stored in such global correlations {could not} be accessed locally. Nonetheless it can again be useful to split the energy {in specific ways (depending on the respective frequencies) between two (or more) batteries, since the optimal protocol brings about a sublinear increase of $(\Delta W)^{2}$ with the input energy.}

\vspace*{-3mm}
\section{Conclusion}\label{sec:discussion}

In this work, we have investigated fundamental and practical limitations on the precision of charging quantum batteries and on the work fluctuations occurring during the charging process. The battery systems we consider are infinite-dimensional bosonic systems, i.e., collections of harmonic oscillators, which are paradigmatic in the theoretical description of physical systems in quantum optics and quantum field theory {and are hence of high conceptual significance}. We assume these systems to be initially thermalized at the ambient temperature. That is, from the point of view of a resource theory of extractable work, empty batteries are considered to be for free, as no work can be extracted from them. We find that, on the one hand, neither the fluctuations nor the precision of the charge for any finite energy input are bounded from above in principle when increasing the average energy of such batteries. On the other hand, we are able to provide lower bounds for both quantities, presenting the respective optimal protocols minimizing the energy variance or fluctuations at given energies and temperatures.\\[-1mm]

In general, these optimal protocols, though theoretically easily describable, are practically difficult to realize, since they require sequences of precise interventions in particular subspaces of the corresponding infinite-dimensional Hilbert spaces. Therefore, it is interesting to understand which limitations apply in scenarios where the energy storage is performed using practically realizable transformation. A set of operations that can usually be implemented comparably simply in such systems is the family of Gaussian unitaries. Here, we have determined the optimal and worst-case Gaussian operations for charging quantum batteries. We find that energy increase via pure single-mode squeezing is the least favourable operation if one wishes to obtain a precise charge for single-mode batteries, whereas the optimal precision, as well as the smallest and largest fluctuations within the restricted set of Gaussian operations are obtained for combinations of squeezing and displacements. For multiple modes, the situation becomes more complicated in principle, but it can be said that it is in general useful to have access to multiple batteries, and that correlations between them are not necessarily detrimental, but also only helpful indirectly.\\[-1mm]

Overall, we conclude that while the optimal Gaussian operations do not achieve results comparable in quality with optimal non-Gaussian protocols, the worst performance achieved with Gaussian operations still produces finite variances and fluctuations, whereas this is not guaranteed in general. In particular, the relative variance and {relative} fluctuations w.r.t. the energy input asymptotically vanish for large energy supply for the optimal Gaussian charging operations. Gaussian unitaries are hence nonetheless practically useful for battery charging. In a sense, Gaussian operations hence represent a trade-off between performance versus reliability and practicality. This is reminiscent of similar contrasts between usefulness and severe limitations of Gaussian operations for tasks in quantum information, e.g., the non-universality of Gaussian operations for quantum computation~\cite{LloydBraunstein1999}. Another observation of this kind can also be made in a different quantum thermodynamical context, where Gaussian operations achieve optimal scaling for the entanglement creation for large energy inputs, but fail to create entanglement in thermal states of finite temperatures when the energy input is too small~\cite{BruschiPerarnauLlobetFriisHovhannisyanHuber2015, BruschiFriisFuentesWeinfurtner2013}.\\[-1mm]

This work hence adds to recent efforts~\cite{BrownFriisHuber2016} of understanding the usefulness of Gaussian operations for quantum thermodynamical tasks, providing investigations of Gaussian unitary work extraction and energy increase. Nonetheless, future work may expand on a number of open questions. For instance, we have here mostly focused on individual batteries since any work stored in joint systems would also require joint extraction. In other words, the role of correlations for work fluctuations and charging precision may be of interest, in particular, in relation to recent results on the work-cost of creating correlations~\cite{BruschiPerarnauLlobetFriisHovhannisyanHuber2015, PerarnauLlobetHovhannisyanHuberSkrzypczykBrunnerAcin2015, HuberPerarnauHovhannisyanSkrzypczykKloecklBrunnerAcin2015, FriisHuberPerarnauLlobet2016, BrunelliGenoniBarbieriPaternostro2017}. In addition, it would also be of interest to consider the consequences of restricting to Gaussian operations for the charging speed (or charging power) as considered in Ref.~\cite{BinderVinjanampathyModiGoold2015, CampaioliPollockBinderCeleriGooldVinjanampathyModi2017}. {Finally, we note that, while some of the results presented here (e.g., the optimal precision charging protocol) directly translate to finite-dimensional systems, other aspects of this work are applicable only to the infinite-dimensional case. An in-depth investigation of the fundamental and practical limitations of precision and fluctuations in charging finite-dimensional systems, although certainly of interest}~\cite{FerraroCampisiAndolinaPellegriniPolini2018}{, goes beyond the scope of this paper.}


\begin{acknowledgments}
We are grateful to Antonio Ac{\'i}n, Eric G. Brown, Nicolas Cerf, and Mart{\'i} Perarnau-Llobet for fruitful discussions and valuable insights. We acknowledge support by the EU COST Action MP1209 ``Thermodynamics in the quantum regime"
and from the Austrian Science Fund (FWF) through the  START  project Y879-N27 and the joint Czech-Austrian project Multi-
QUEST (I 3053-N27).
\end{acknowledgments}


\hypertarget{sec:appendix}
\appendix
\section*{Appendix}
\renewcommand{\thesubsubsection}{A.\arabic{subsection}.\Roman{subsubsection}}
\renewcommand{\thesubsection}{A.\arabic{subsection}}
\renewcommand{\thesection}{A}
\setcounter{equation}{0}
\numberwithin{equation}{section}
\setcounter{figure}{0}
\renewcommand{\thefigure}{A.\arabic{figure}}

\subsection{Optimal precision charging protocol}\label{sec:optimal general charging protocol}

In this appendix, we give a detailed description of a unitary battery charging protocol that raises the average energy of an initial single-mode thermal state whilst keeping the energy variance of the final state minimal. The initial thermal state with density operator $\tau(\beta)$ is diagonal in the energy basis. The corresponding diagonal elements are the probability weights $p_{n}=(1-e^{-\beta\omega})e^{-n\beta\omega}$, which are decreasing with increasing energies $E_{n}=n\omega$. The initial average energy $E\bigl(\tau(\beta)\bigr)=\epsilon_{o}\omega$, where $\epsilon_{o}=e^{-\beta\omega}(1-e^{-\beta\omega})^{-1}$ and the initial energy variance $V\bigl(\tau(\beta)\bigr)=\omega^{2}e^{-\beta\omega}(1-e^{-\beta\omega})^{-2}$ are determined by the initial temperature $T=1/\beta$. We are then interested in increasing the average energy by an amount $\Delta E=\omega\Delta\epsilon$ to reach a state $\rho$ with $E(\rho)=\omega\epsilon=E(\tau)+\Delta E$. In particular, we aim to achieve this increase unitarily, i.e., such that $\rho=U\tau U^{\dagger}$. Moreover, we want to keep the energy variance of $\rho$ minimal. In other words, we would like to determine the (non-unique) minimal energy-variance state $\rho$ with given average energy $\omega\epsilon$ in the unitary orbit of $\tau(\beta)$.

{The protocol that we present here to obtain such a state consists of two parts} (\hyperref[sec:part I of protocol]{I} \& \hyperref[sec:part II of protocol]{I\nl I}). {Each of these parts can be described as a series of (unitary) two-level rotations, ensuring that the final state is within the unitary orbit of the initial state. The two-level rotations between pairs of energy levels are used to appropriately shift and reorder the probability weights $p_{n}$ of the initial state. As we shall explain, part}~\hyperref[sec:part I of protocol]{I} {of the protocol reaches the unique state $\tilde{\rho}$ in the unitary orbit of $\tau(\beta)$ that minimizes the average squared distance to the target energy. The state $\tilde{\rho}$ is diagonal in the energy eigenbasis, and arises from a permutation of the weights $p_{n}$ that assigns positions with increasing distance to the target energy to weights with decreasing size. However, the state obtained in this way does not have the desired target energy, i.e., in general $E(\tilde{\rho})\neq E(\rho)$. During part}~\hyperref[sec:part II of protocol]{I\nl I} {of the protocol, this deviation of the average energy is corrected in such a way that the target energy is reached whilst only minimally increasing the average squared distance to it.}

\vspace*{-3mm}
\subsubsection{Part~I of the protocol}\label{sec:part I of protocol}
\vspace*{-1mm}

{In part}~\hyperref[sec:part I of protocol]{I}, we first identify the energy level (labelled $k$) that is closest to the desired target energy, i.e., we define
\vspace*{-1mm}
\begin{align}
    k   &=
    \begin{cases}
        \lfloor \epsilon \rfloor    & \mbox{if}\ \ \epsilon-\lfloor\epsilon\rfloor<\lceil\epsilon\rceil-\epsilon\\[1mm]
        \lceil \epsilon \rceil    & \mbox{if}\ \ \epsilon-\lfloor\epsilon\rfloor\geq\lceil\epsilon\rceil-\epsilon
    \end{cases},
\end{align}
where we distinguish between two cases, depending on whether $\epsilon$ is closer to the energy level above or below its value. The probability weights for the case where $k=\lfloor \epsilon \rfloor$ are illustrated in Fig.~\ref{fig:charging protocol start}~(a). Part~\hyperref[sec:part I of protocol]{I} of the protocol then consists of a reordering of the weights $p_{n}$ such that the largest weight $p_{0}$ is moved to the energy level $k$, the second largest weight $p_{1}$ to the second closest level to $\epsilon$, and so forth. After part~\hyperref[sec:part I of protocol]{I}, the density operator is still diagonal, but the probability weights on the diagonal are now either given by
\begin{align}
    \tilde{p}_{n}   &=\,
    \begin{cases}
        p_{2(k-n)}  & \mbox{for}\ \ n=0,\ldots,k\\[1mm]
        p_{2(n-k)-1}  & \mbox{for}\ \ n=k+1,\ldots,\max\{1,2k\}\\[1mm]
        p_{n}  & \mbox{for}\ \ n\geq\max\{2,2k+1\}
    \end{cases},
\end{align}
if $k=\lfloor \epsilon \rfloor$, or, in case that $k=\lceil \epsilon \rceil$ by
\vspace*{-1mm}
\begin{align}
    \tilde{p}_{n}   &=\,
    \begin{cases}
        p_{2(k-n)-1}  & \mbox{for}\ \ n=0,\ldots,k-1\\[1mm]
        p_{2(n-k)}  & \mbox{for}\ \ n=k,\ldots,2k-1\\[1mm]
        p_{n}  & \mbox{for}\ \ n\geq2k
    \end{cases}.
\end{align}

The resulting probability distribution, illustrated in Fig.~\ref{fig:charging protocol start}~(b) for the case $k=\lfloor \epsilon \rfloor$, has an average energy $\tilde{\epsilon}_{\mathrm{I}}=\sum_{n}\tilde{p}_{n}n$ and its average squared distance from the target $\epsilon$ is minimal, i.e., we have arrived at the unique state $\tilde{\rho}$ in the unitary orbit of the initial state $\tau$ that minimizes $\tilde{V}_{\mathrm{I}}=\sum_{n}\tilde{p}_{n}(n-\epsilon)^{2}$. However, as  $\tilde{\epsilon}_{\mathrm{I}}$ in general does not match $\epsilon$, which implies that $\tilde{V}_{\mathrm{I}}$ also is not equal to the energy variance, we are not yet done. Interestingly, for both $k=\lfloor\epsilon\rfloor$ and $k=\lceil\epsilon\rceil$ one may encounter combinations of $T$ and $\Delta\epsilon$ such that $\tilde{\epsilon}_{\mathrm{I}}<\epsilon$ or $\tilde{\epsilon}_{\mathrm{I}}>\epsilon$.

\subsubsection{Part~I\nl I of the protocol}\label{sec:part II of protocol}

In part~\hyperref[sec:part II of protocol]{I\nl I} of the protocol we hence have to appropriately adjust the average energy. This can again be done by a sequence of two-level rotations. Each of this transformations will bring the average energy closer to $\epsilon$, but since we start from a minimum of the average squared deviation from $\epsilon$, the value of the latter will increase. We are hence interested in selecting the optimal sequence of these two-level rotations. To start, consider a rotation between the levels $m$ and $n$ with weights $\tilde{p}_{m}$ and $\tilde{p}_{n}$ by an angle $\theta$. This corresponds to the map
\begin{align}
    (\tilde{p}_{m},\tilde{p}_{n})   &\mapsto(\cos^{2\!}\!\theta\,\tilde{p}_{m}+\sin^{2\!}\!\theta\, \tilde{p}_{n},\cos^{2\!}\!\theta\,\tilde{p}_{n}+\sin^{2\!}\!\theta\,\tilde{p}_{m}),
\end{align}
and leads to a change in energy given by
\begin{align}
    \Delta\tilde{\epsilon}    &=\,\sin^{2\!}\!\theta\,(\tilde{p}_{n}-\tilde{p}_{m})(m-n).
    \label{eq:energy increase opt prot general}
\end{align}
Meanwhile, the increase of the mean squared deviation from $\epsilon$ can be written as
\begin{align}
    \Delta\tilde{V} &=\omega^{2}\sin^{2\!}\!\theta\,(\tilde{p}_{n}-\tilde{p}_{m})\bigl((m-\epsilon)^{2}-(n-\epsilon)^{2}\bigr).
\end{align}

\vspace*{-4mm}
\clearpage
\begin{figure}[ht!]
\label{fig:charging protocol start}
\begin{center}
(a)\includegraphics[width=0.41\textwidth,trim={0cm 0mm 0cm 0mm}]{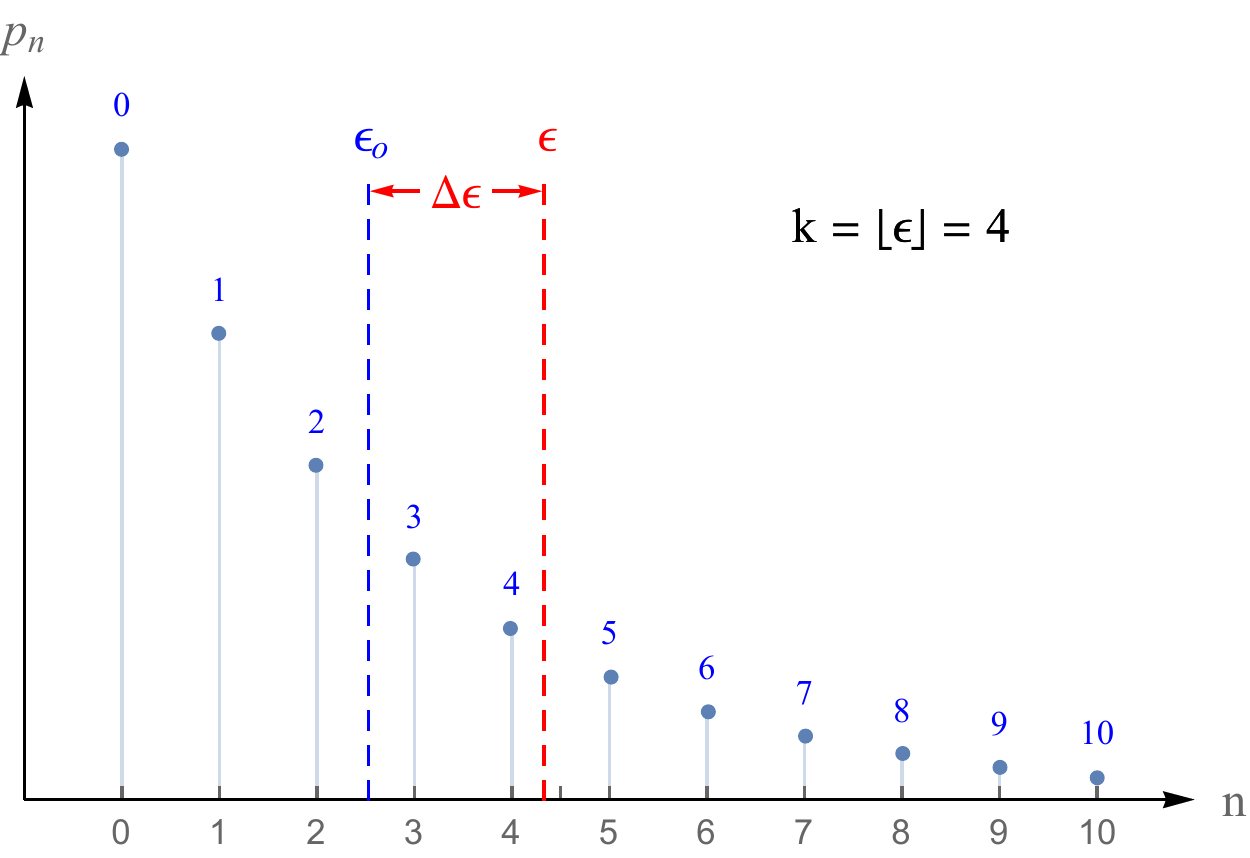}
(b)\includegraphics[width=0.41\textwidth,trim={0cm 0mm 0cm 0mm}]{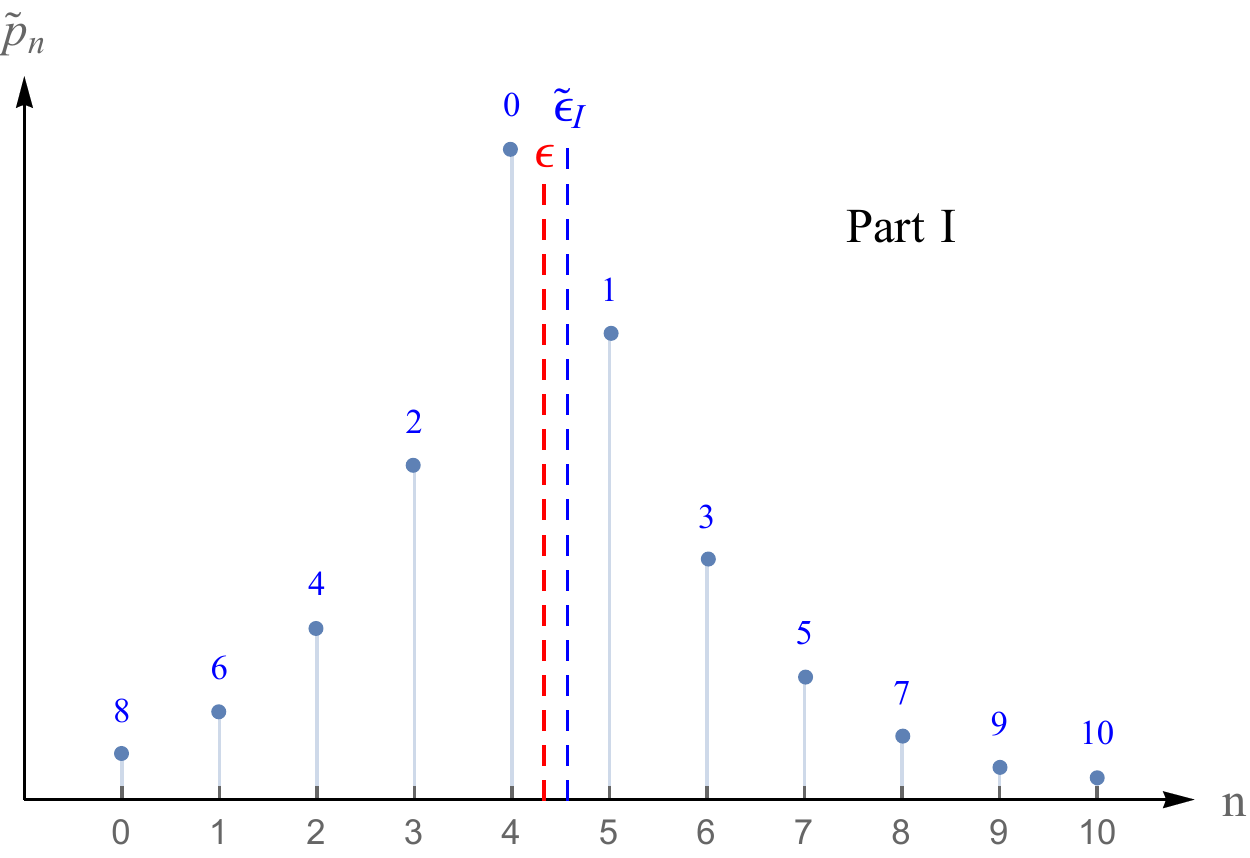}
(c)\includegraphics[width=0.41\textwidth,trim={0cm 0mm 0cm 0mm}]{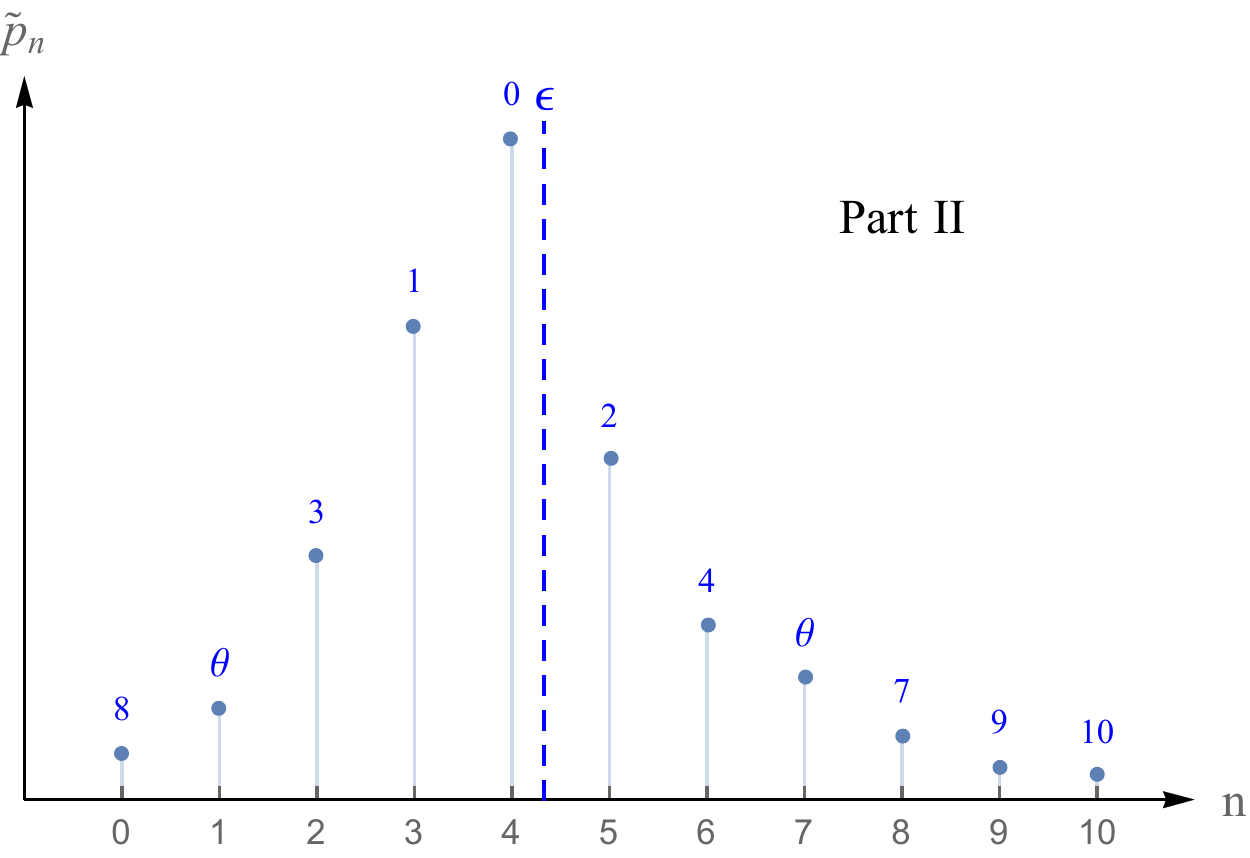}
\end{center}
\vspace*{-3mm}
\caption{
\textbf{Optimal precision battery charging:} The protocol for optimal precision battery charging is illustrated for an initial thermal state of temperature $T=3$ (in units of $\omega$). (a) The probability weights $p_{n}$ of the initial state decrease with increasing energy. The initial average energy $\epsilon_{o}$ (we use the dimensionless variables here for simplicity) is to be raised by $\Delta\epsilon$ to a value $\epsilon$, that is closer to the energy level $k=\lfloor\epsilon\rfloor=4$ rather than $\lceil\epsilon\rceil=5$. (b) Part~\protect\hyperref[sec:part I of protocol]{I}: After rearranging the probability weights to place the largest weights closest to $k$, one obtains a distribution $\{\tilde{p}_{n}\}$ with an average energy of $\tilde{\epsilon}_{\mathrm{I}}>\epsilon$. The numbers $m=0,\ldots,10$ above the vertical lines at horizontal position $n$ indicate that the corresponding weight $\tilde{p}_{n}$ corresponds to the value of the original weight $p_{m}$. (c) Part~\hyperref[sec:part II of protocol]{I\nl I}: Additional two-level rotations adjust the energy to the target $\epsilon$. The first two of these rotations (corresponding to $\varphi=1$ with $j=0$ and $l=1,2$, see Sec.~\ref{sec:part II of protocol}) have angles $\tfrac{\pi}{2}$ and hence completely exchange the populations of the levels $(m,n)=(3,5)$ and $(2,6)$. The third rotation between levels $1$ and $7$ requires only a smaller angle $0<\theta<\tfrac{\pi}{2}$ until reaching the target energy.}
\end{figure}

Now, let us pick two such values on either side of $\epsilon$, i.e., $m=k-l$ and $n=k+l+j$, where $l\in\mathbb{N}_{0}$ determines the distance from the level $k$, and $j\in\mathbb{Z}$ quantifies the difference in distances to $k$ (or equivalently to $\epsilon$) between the levels $m$ and $n$. More specifically, we have
\begin{align}
    d(n,\epsilon)-d(m,\epsilon) &=(n-\epsilon)-(\epsilon-m)=2(k-\epsilon)+j.
\end{align}
Moreover, this implies that the energy change from Eq.~(\ref{eq:energy increase opt prot general}) is given by
\begin{align}
    \Delta\tilde{\epsilon}  &=\,\sin^{2\!}\!\theta\,(\tilde{p}_{n}-\tilde{p}_{m})(2l+j).
\end{align}
With this we further find that the changes of the energy and of $\tilde{V}$ have the relation
\begin{align}
    \tfrac{1}{\omega^{2}}\tfrac{\Delta\tilde{V}}{\Delta\tilde{\epsilon}} &=\,2(k-\epsilon)+j.
\end{align}
With this knowledge, we come to a more detailed
\makebox[\linewidth][s]{description of the protocol. First, we set $\tilde{\epsilon}=\tilde{\epsilon}_{\mathrm{I}}$ and} $\tilde{V}=\tilde{V}_{\mathrm{I}}$, and we distinguish between the two situations $\tilde{\epsilon}<\epsilon$ and $\tilde{\epsilon}>\epsilon$. On the one hand, when $\tilde{\epsilon}<\epsilon$, we need to increase the energy, which means picking levels $m$ and $n$ such that $\tilde{p}_{m}>\tilde{p}_{n}$ and $2l+j>0$, whilst choosing $j$ as small as possible to minimize $\tfrac{\Delta\tilde{V}}{\Delta\tilde{\epsilon}}$. On the other hand, when $\tilde{\epsilon}>\epsilon$, we need to decrease the energy, suggesting that one should select levels $m$ and $n$ such that $\tilde{p}_{m}<\tilde{p}_{n}$ and $2l+j>0$, whilst choosing $j$ as large as possible to minimize $\tfrac{\Delta\tilde{V}}{\Delta\tilde{\epsilon}}$. When such levels are chosen and the potential energy change exceeds what is needed to reach the target, one appropriately fixes the rotation angle $\theta$ such that $\tilde{\epsilon}+\Delta\tilde{\epsilon}=\epsilon$. If the energy change achievable with a specific such rotation is not sufficient to reach the target, one rotates by $\theta=\tfrac{\pi}{2}$, updates $\tilde{\epsilon}$, $\tilde{V}$, and $\{\tilde{p}_{n}\}$ and continues with the next viable pair of levels minimizing $\tfrac{\Delta\tilde{V}}{\Delta\tilde{\epsilon}}$. Inspection of all cases then reveals that the (first phase) of part~\hyperref[sec:part II of protocol]{I\nl I} consists of $k$ or $k+1$ two-level rotations labelled by $l=l_{\mathrm{min}},\ldots,k$, where
\begin{align}
    l_{\mathrm{min}}    &=
    \begin{cases}
        0   &\mbox{if}\ \ k=\lfloor\epsilon\rfloor, \tilde{\epsilon}<\epsilon\\[1mm]
        1   &\mbox{otherwise}
    \end{cases},
\end{align}
and for each of these rotations we choose
\begin{align}
    j    &=
    \begin{cases}
        +1 &\mbox{if}\ \ k=\lfloor\epsilon\rfloor, \tilde{\epsilon}<\epsilon\\[1mm]
        \ 0 &\mbox{if}\ \ k=\lfloor\epsilon\rfloor, \tilde{\epsilon}>\epsilon\\[1mm]
        \ 0 &\mbox{if}\ \ k=\lceil\epsilon\rceil, \tilde{\epsilon}<\epsilon\\[1mm]
        -1 &\mbox{if}\ \ k=\lceil\epsilon\rceil, \tilde{\epsilon}>\epsilon
    \end{cases}.
\end{align}
Since $\tfrac{\Delta\tilde{V}}{\Delta\tilde{\epsilon}}$ does not depend on $l$ and the rotations all commute (they pertain to different subspaces), the order of these operations within the first phase is irrelevant. However, not even all $k$ (or $k+1$) rotations may
\clearpage

\noindent generally be enough to sufficiently adjust the average energy. Consequently, part~\hyperref[sec:part II of protocol]{I\nl I} may consist of an arbitrary number of phases labelled by $\varphi=1,2,\ldots$, where
\begin{align}
    \bigl(j(\varphi),l_{\mathrm{min}}(\varphi)\bigr)    &=
    \begin{cases}
        (\varphi,-\lceil\tfrac{\varphi}{2}\rceil+1) \!\!\!&\!\mbox{if}\ k=\lfloor\epsilon\rfloor, \tilde{\epsilon}<\epsilon\\[1mm]
        (-\varphi+1,\lceil\tfrac{\varphi}{2}\rceil) \!\!\!&\!\mbox{if}\ k=\lfloor\epsilon\rfloor, \tilde{\epsilon}>\epsilon\\[1mm]
        (\varphi\!-\!1,-\lfloor\tfrac{\varphi}{2}\rfloor\!+\!1) \!\!\!&\!\mbox{if}\ k=\lceil\epsilon\rceil, \tilde{\epsilon}<\epsilon\\[1mm]
        (-\varphi,\lfloor\tfrac{\varphi}{2}\rfloor+1) \!\!\!&\!\mbox{if}\ k=\lceil\epsilon\rceil, \tilde{\epsilon}>\epsilon
    \end{cases}.
\end{align}

Let us now give a more compact description of part~\hyperref[sec:part II of protocol]{I\nl I}. After part~$\mathrm{I}$, set $\{\tilde{p}\}$ as the initial distribution, and further set $\tilde{\epsilon}=\tilde{\epsilon}_{\mathrm{I}}$, $\tilde{V}=\tilde{V}_{\mathrm{I}}$, and $\varphi=1$. Then perform the following steps:

\begin{enumerate}[(i)]
\item{\label{eq:protocol step i}Set $j=j(\varphi)$, and $l=l_{\mathrm{min}}(\varphi)$.}
\item{\label{eq:protocol step ii}If $\tilde{\epsilon}\neq\epsilon$ and $l\leq k$, set $m=k-l$, $n=k+l+j$, $\Delta\tilde{\epsilon}_{\mathrm{max}}\suptiny{1}{0}{\textrm{I\nl I}}=(\tilde{p}_{m}-\tilde{p}_{n})(2l+j)$, and continue with step~(\ref{eq:protocol step iii}). If $\tilde{\epsilon}\neq\epsilon$ and $l>k$, increase $\varphi$ by one, i.e., $\varphi\mapsto\varphi+1$ and start again with step~(\ref{eq:protocol step i}). If $\tilde{\epsilon}=\epsilon$ the protocol concludes.}
\item{\label{eq:protocol step iii}If $\tilde{\epsilon}<\epsilon$, then $\Delta\tilde{\epsilon}_{\mathrm{max}}\suptiny{1}{0}{\textrm{I\nl I}}>0$ and $\theta_{l}$ is set to the value
\begin{align}
    \theta_{l}  &=\,\begin{cases}
        \tfrac{\pi}{2}  & \mbox{if}\ \ \tilde{\epsilon}+\Delta\tilde{\epsilon}_{\mathrm{max}}\suptiny{1}{0}{\textrm{I\nl I}}<\epsilon\\[1mm]
        \arcsin\sqrt{\tfrac{\epsilon-\tilde{\epsilon}}{\Delta\tilde{\epsilon}_{\mathrm{max}}\suptiny{1}{0}{\textrm{I\nl I}}}}
        & \mbox{if}\ \ \tilde{\epsilon}+\Delta\tilde{\epsilon}_{\mathrm{max}}\suptiny{1}{0}{\textrm{I\nl I}}\geq\epsilon
    \end{cases}.
\end{align}
If $\tilde{\epsilon}>\epsilon$, then $\Delta\tilde{\epsilon}_{\mathrm{max}}\suptiny{1}{0}{\textrm{I\nl I}}<0$ and $\theta_{l}$ is set to the value
\begin{align}
    \theta_{l}  &=\,\begin{cases}
        \tfrac{\pi}{2}  & \mbox{if}\ \ \tilde{\epsilon}+\Delta\tilde{\epsilon}_{\mathrm{max}}\suptiny{1}{0}{\textrm{I\nl I}}>\epsilon\\[1mm]
        \arcsin\sqrt{\tfrac{\epsilon-\tilde{\epsilon}}{\Delta\tilde{\epsilon}_{\mathrm{max}}\suptiny{1}{0}{\textrm{I\nl I}}}}
        & \mbox{if}\ \ \tilde{\epsilon}+\Delta\tilde{\epsilon}_{\mathrm{max}}\suptiny{1}{0}{\textrm{I\nl I}}\leq\epsilon
    \end{cases}.
\end{align}
Then continue with step~(\ref{eq:protocol step iv}).}
\item{\label{eq:protocol step iv}Perform the following updates:
    \begin{subequations}
    \begin{align}
        \tilde{p}_{m} &\ \mapsto \cos^{2\!}\!\theta_{l}\,\tilde{p}_{m}+\sin^{2\!}\!\theta_{l}\, \tilde{p}_{n},\nonumber\\[1mm]
        \tilde{p}_{n} &\ \mapsto \cos^{2\!}\!\theta_{l}\,\tilde{p}_{n}+\sin^{2\!}\!\theta_{l}\,\tilde{p}_{m},\nonumber\\[1mm]
        \tilde{\epsilon}    &\ \mapsto\ \tilde{\epsilon}+\sin^{2\!}\!\theta_{l}\,\Delta\tilde{\epsilon}_{\mathrm{max}}\suptiny{1}{0}{\textrm{I\nl I}}=\Delta\tilde{\epsilon}\suptiny{1}{0}{\textrm{I\nl I}},\nonumber\\[1mm]
        \tilde{V} &\ \mapsto\ \tilde{V}+\Delta\tilde{\epsilon}\suptiny{1}{0}{\textrm{I\nl I}}(2(k-\epsilon)+j),\nonumber
    \end{align}
    \end{subequations}
    Finally, increase $l$ by one and start over from step~(\ref{eq:protocol step ii}).}
\end{enumerate}
After the conclusion of part~\hyperref[sec:part II of protocol]{I\nl I}, the target energy has been reached, $\tilde{\epsilon}=\epsilon$, and the average squared deviation from $\epsilon$ hence becomes the energy variance. The second part of the protocol is illustrated in Fig.~\ref{fig:charging protocol start}~(c) and the variances resulting from the protocol for different temperatures and input energies are shown in Fig.~\ref{fig:optimal charging} of the main text.

\subsection{Wigner representation of squared number operator}\label{sec:wigner rep of squared number op}

We do this by using the formulas of Eqs.~(\ref{eq:exp value via Wigner function}) and~(\ref{eq:Wigner function Gaussian states}). To this end, we start by rewriting {$\hat{N}^{2}$} in terms of the local position and momentum operators as
\begin{align}
    \hat{N}^{2} &=\,\tfrac{1}{4}\bigl(\hat{x}^{2}+\hat{p}^{2}-1\bigr)^{2}\\[1mm]
    &=\,\tfrac{1}{4}\bigl(\hat{x}^{4}+\hat{p}^{4}+\hat{x}^{2}\nr\hat{p}^{2}+\hat{p}^{2}\nr\hat{x}^{2}-2(\hat{x}^{2}+\hat{p}^{2})-1\bigr).
    \nonumber
\end{align}
We then insert term by term into Eq.~(\ref{eq:Wigner transform}) and calculate
\begin{align}
    &\bra{x-\tfrac{y}{2}}f(\hat{x})\ket{x+\tfrac{y}{2}}\nonumber\\[1mm]
    &\ \hspace*{8mm} =\bra{x-\tfrac{y}{2}}f(x\!+\!\tfrac{y}{2})\ket{x+\tfrac{y}{2}}\nonumber\\[1mm]
    &\ \hspace*{8mm} =f(x+\tfrac{y}{2})\scpr{x-\tfrac{y}{2}}{x+\tfrac{y}{2}}\nonumber\\[1mm]
    &\ \hspace*{8mm} =f(x+\tfrac{y}{2})\,\delta(y),
\end{align}
\begin{align}
    &\bra{x-\tfrac{y}{2}}f(\hat{p})\ket{x+\tfrac{y}{2}}\nonumber\\[1mm]
    &\ \hspace*{8mm} =\,\bra{x-\tfrac{y}{2}}f(\hat{p})\!\!\int\!\!dp\pr\ket{p\pr}\!\scpr{p\pr}{x+\tfrac{y}{2}}\nonumber\\[1mm]
    &\ \hspace*{8mm} =\!\int\!\!dp\pr\,f(p\pr)\scpr{x-\tfrac{y}{2}}{p\pr}\!\scpr{p\pr}{x+\tfrac{y}{2}}\nonumber\\[1mm]
    &\ \hspace*{8mm} =\,\tfrac{1}{2\pi}\!\int\!\!dp\pr\,f(p\pr)\,e^{-i\nr p\pr y},
\end{align}
for functions $f$ of the operators {$\hat{x}$} and {$\hat{p}$}, where we have used that
\begin{align}
    \scpr{x}{x\pr}  &=\,\tfrac{1}{2\pi}\int\!\!dp\,e^{i\nr p\nr(x-x\pr)}\,=\,\delta(x-x\pr)\,,
    \label{eq:overlap x with xpr}\\[1mm]
    \scpr{x}{p}  &=\,\tfrac{1}{(2\pi)^{1/2}}\,e^{i\nr p\nr x}\,.
    \label{eq:overlap x with p}
\end{align}
After some algebra we then find the phase space representation of the operator {$\hat{N}^{2}$} to be given by
\begin{align}
    N^{2}(x,p)  &=\tfrac{1}{4}\bigl(x^{2}+p^{2}-1\bigr)^{2}\nonumber\\[1mm]
    & \ \ +\tfrac{1}{16\pi}\!\!\int\!\!d\tilde{q}d\tilde{p}\tilde{q}^{2}\tilde{p}^{2} e^{i(p-\tilde{p})\tilde{q}}.
    \label{eq:Wigner transform of N squared full}
\end{align}

The second term on the right-hand side can be understood in the distributional sense. That is, defining the distribution {$\gamma[g(p)]$} via the function {$\gamma(p)$} given by
\begin{align}
    \gamma(p)  &:=\,\int\!\!dq\,q^{2}\,e^{-2\nr i\nr pq}
    \label{eq:gamma functional definition}
\end{align}
one finds that for any Schwartz function {$g(p)$} we have
\begin{align}
    \gamma[g(p)]    &=\!\!\int\!\!dp\gamma(p)\,g(p)\,
    =\,-\,\tfrac{\pi}{4}\left.\tfrac{\partial^{2}}{\partial p^{2}}g(p)\right|_{p=0}\nonumber\\[1mm]
    &\ =\,-\,\tfrac{\pi}{4}\,g\prpr(0).
    \label{eq:gamma functional}
\end{align}
Then note that the wave function {$\psi(x)=\scpr{x}{\psi}$} of every single-mode pure state $\ket{\psi}$ can be expanded in terms of the Hermite polynomials {$H_{j}(x)$} as
\begin{align}
    \psi(x) &=\,\tfrac{1}{\pi^{1/4}}e^{-x^{2}/2}\sum\limits_{j}\tfrac{c_{j}}{\sqrt{2^{j}j!}}H_{j}(x).
    \label{eq:general single mode wave function}
\end{align}
with $\sum_{j}|c_{j}|^{2}=1$. Using Eq.~(\ref{eq:overlap x with xpr}) we can then write the Wigner function for an arbitrary single-mode pure {state as}
\begin{align}
    \mathcal{W}(x,p)    &=\,\tfrac{1}{\pi^{3/2}}e^{-x^{2}}\!\!\int\!\!dy\,e^{-y^{2}-2\nr i\nr py}\,h(x,y)\,
\end{align}
with the function
\begin{align}
    h(x,y)&=\sum\limits_{j,k}\tfrac{c_{j}c_{k}^{*}}{\sqrt{2^{j+k}j!k!}}H_{j}(x+y)H^{*}_{k}(x-y).\nonumber
\end{align}
Finally, we can compute the integral of {$\mathcal{W}(x,p)$} with the second term on the right-hand side of Eq.~(\ref{eq:Wigner transform of N squared full}) and find
\begin{align}
    &\tfrac{1}{16\pi}\!\!\int\!\!dx\,dp\,\mathcal{W}(x,p)\!\!\int\!\!d\tilde{q}\,d\tilde{p}\,\tilde{q}^{2}\tilde{p}^{2} e^{i(p-\tilde{p})\tilde{q}}\nonumber\\[1mm]
    &=\tfrac{1}{16\pi^{5/2}}\!\!\int\!\!dxe^{-x^{2}}\!\!\int\!\!dy e^{-y^{2}}h(x,y)\nonumber\\[1mm]
    &\ \times\!\!\int\!\!d\tilde{p}\tilde{p}^{2}\!\!\int\!\!d\tilde{q}\tilde{q}^{2}e^{-i\tilde{p}\tilde{q}}
    \!\!\int\!\!dp e^{ip(\tilde{q}-2y)}\nonumber\\[1mm]
    &=-\tfrac{1}{8\pi^{1/2}}\!\!\int\!\!dx e^{-x^{2}}\left.\tfrac{\partial^{2}}{\partial y^{2}}\bigl(y^{2}e^{-y^{2}}h(x,y)\bigr)\right|_{y=0}.
    \nonumber
\end{align}
where we have integrate over {$p$} using Eq.~(\ref{eq:overlap x with xpr}), followed by an integral over the delta function {$\delta(\tilde{q}-2y)$}, and finally made use of Eqs.~(\ref{eq:gamma functional definition}) and~(\ref{eq:gamma functional}). It is then easy to see that only the term {$2h(x,0)$} remains after taking the derivatives and evaluating at {$y=0$}. Using the normalization of the wave function in Eq.~(\ref{eq:general single mode wave function}) and arrive at
\begin{align}
    \tfrac{1}{4\pi^{1/2}}\!\!\int\!\!dx e^{-x^{2}}h(x,0)    &=\tfrac{1}{4}\!\!\int\!\!dx\,\psi(x)\psi^{*}(x)=\tfrac{1}{4}.
    \nonumber
\end{align}
Due to the linearity of the Wigner function in $\rho$, this computation extends from $\ket{\psi}$ to arbitrary single-mode mixed states, and since {the number operator of each mode} is a local observable also to arbitrary $N$-mode states. We can hence rewrite Eq.~(\ref{eq:Wigner transform of N squared full}) as
\begin{align}
    N^{2}(x,p)  &=\,\tfrac{1}{4}\bigl(x^{2}+p^{2}-1\bigr)^{2}\,-\,\tfrac{1}{4}.
    \label{eq:Wigner transform of N squared simplified}
\end{align}

\subsection{Extremal Gaussian precision and fluctuations}\label{sec:Extremal Gaussian Precision and Fluctuations}

In this appendix, we give detailed proofs for the existence and uniqueness of the extremal values of the charging precision and fluctuations when restricting to single-mode Gaussian unitaries at fixed initial temperature and energy input. The corresponding results are presented and discussed in Sections~\ref{sec:Optimal and worst-case Gaussian charging precision} and~\ref{sec:Charging fluctuations for single-mode Gaussian unitaries}.

\subsubsection{Maximal variance}\label{sec:max variance}

We begin with the maximally possible variance that single-mode Gaussian unitaries allow for, i.e., the worst-case scenario. Here, one is interested in determining the maximum of the function $V_{+}(r)$ from Eq.~(\ref{eq:VmaxVmin}) over all $r\geq0$. For brevity, we will (again) use the notation $\Delta\epsilon=\Delta E/\omega$ and $\nu=\coth\bigl(\tfrac{\beta\omega}{2}\bigr)$, as well as define $\mathcal{V}_{+}:=4V_{+}/\omega^{2}$, such that the function that we wish to maximize can be written as
\begin{align}
    \mathcal{V}_{+} &=\,\nu^{2}\cosh(4r)-1+2\nu e^{2r}\bigl( 2\Delta\epsilon-\nu\bigl[\cosh(2r)-1\bigr] \bigr)
    \nonumber\\
    &=\,-\nu^{2}\bigl[\sinh(4r)-2e^{2r}(2\tfrac{\Delta\epsilon}{\nu}+1)\bigr]-(\nu^{2}+1).
    \label{eq:Vmax appendix}
\end{align}
To determine the extremal points, we calculate the first and second partial derivatives of $\mathcal{V}_{+}$ w.r.t. $r$, i.e.,
\begin{align}
    \frac{\partial\mathcal{V}_{+}}{\partial r}  &=\,
    -4\nu^{2}\bigl[\cosh(4r)-(2\tfrac{\Delta\epsilon}{\nu}+1)e^{2r}\bigr],
    \label{eq:partial deriv of V+}\\[1mm]
    \frac{\partial^{2}\mathcal{V}_{+}}{\partial r^{2}}  &=\,
    -8\nu^{2}\bigl[2\sinh(4r)-(2\tfrac{\Delta\epsilon}{\nu}+1)e^{2r}\bigr].
    \label{eq:second partial deriv of V+}
\end{align}
The condition $\tfrac{\partial\mathcal{V}_{+}}{\partial r}|_{r=r_{\mathrm{extr.}}}=0$ at the extremal points $r=r_{\mathrm{extr.}}$ yields
\begin{align}
    (2\tfrac{\Delta\epsilon}{\nu}+1)  &=\,e^{-2r_{\mathrm{extr.}}}\,\cosh(4r_{\mathrm{extr.}})\,.
    \label{eq:prec worst case extr condition}
\end{align}
Now note that the left-hand side is greater or equal than $1$, while the function $f(r)=e^{-2r}\cosh(4r)$ appearing on the right-hand side satisfies $f(r=0)=1$ and has a minimum at $r=\ln(3)/8>0$ (as can be seen by setting $\partial f(r)/\partial r=0$), and $f(r)\substack{\text{\tiny{$r\!\rightarrow\!\infty$}}\\ \longrightarrow}\infty$. Consequently, Eq.~(\ref{eq:prec worst case extr condition}) has two solutions $r_{\mathrm{extr.}}^{\pm}$, with $r_{\mathrm{extr.}}^{+}>\ln(3)/8>0$ and $r_{\mathrm{extr.}}^{-}<0$. Inserting $(2\tfrac{\Delta\epsilon}{\nu}+1)$ from Eq.~(\ref{eq:prec worst case extr condition}) back into the second partial derivative gives
\begin{align}
    \left.\frac{\partial^{2}\mathcal{V}_{+}}{\partial r^{2}}\right|_{r=r_{\mathrm{extr.}}}  &=\,
    -4\nu^{2}\bigl[e^{4r_{\mathrm{extr.}}}-3e^{-4r_{\mathrm{extr.}}}\bigr],
\end{align}
which is negative for $r_{\mathrm{extr.}}^{+}$ (which is $>\ln(3)/8$) and positive for $r_{\mathrm{extr.}}^{-}<0$. We thus have a local maximum at $r=r_{\mathrm{extr.}}^{+}$, while $r_{\mathrm{extr.}}^{-}$ is a local minimum. Then, recall that we are interested in non-negative solutions (a negative squeezing parameter would reverse the roles of $\xi_{1}$ and $\xi_{2}$ in our treatment, see Section~\ref{sec:Optimal and worst-case Gaussian charging precision}), and can thus discard $r_{\mathrm{extr.}}^{-}$. Moreover, this eliminates all values of $\mathcal{V}_{+}(r)$ for $r<r_{\mathrm{extr.}}^{-}$, which can become larger than $\mathcal{V}_{+}(r_{\mathrm{extr.}}^{+})$. In other words, in the relevant range $r\geq0$, a (global) maximum of $\mathcal{V}_{+}(r)$ can be found at $r_{\mathrm{extr.}}^{+}$. Nevertheless, this is not the sought-after maximum for the variance, as we shall explain next.

If $r_{\mathrm{extr.}}^{+}>0$, implicitly determined by Eq.~(\ref{eq:prec worst case extr condition}), were the correct solution, we could express the total input energy as
\begin{align}
    \Delta\epsilon  &=\,\tfrac{\nu}{2}\bigl(e^{-2r_{\mathrm{extr.}}^{+}}\,\cosh(4r_{\mathrm{extr.}}^{+})-1\bigr)\,.
\end{align}
Since the energy (in units of $\omega$) invested into squeezing as a function of the squeezing parameter is given by $\Delta\epsilon_{\mathrm{sq}}=\tfrac{\nu}{2}\bigl(\cosh(2r)-1\bigr)$, see Eq.~(\ref{eq:Delta E for single mode Gaussian states}), we could then write the energy invested into squeezing for $r=r_{\mathrm{extr.}}^{+}$ as
\begin{align}
    \bigl(\Delta\epsilon_{\mathrm{d}}\bigr)_{r=r_{\mathrm{extr.}}^{+}} &=\,\bigl(\Delta\epsilon-\Delta\epsilon_{\mathrm{sq}}\bigr)_{r=r_{\mathrm{extr.}}^{+}}\\
    &=\,\tfrac{\nu}{4}\bigl(e^{-6r_{\mathrm{extr.}}^{+}}-e^{-2r_{\mathrm{extr.}}^{+}}\bigr)<0.\nonumber
\end{align}
In other words, the (local) maximum of the function $\mathcal{V}_{+}(r)$ at $r=r_{\mathrm{extr.}}^{+}$ is not physically realizable, since we must require $\Delta\epsilon_{\mathrm{d}}\geq0$. Put simply, the maximum at $r=r_{\mathrm{extr.}}^{+}$ would require more energy to be invested into squeezing than is available overall,  $\Delta\epsilon_{\mathrm{sq}}>\Delta\epsilon$. Since $\Delta\epsilon_{\mathrm{sq}}$ is strictly increasing with increasing squeezing parameter, we are hence looking for a solution for some $r=\rplus<r_{\mathrm{extr.}}^{+}$ that maximizes $\mathcal{V}_{+}$ within the physically allowed range. Our previous analysis informs us that such a solution exists uniquely. The function $\mathcal{V}_{+}(r)$ has one local minimum at a negative argument, and one local maximum for $r>0$, and must hence be strictly increasing between $r$ and $r_{\mathrm{extr.}}^{+}$. The solution we are looking for is thus unique and found for the maximal value $r=\rplus$ allowed by the global energy constraint, that is
\begin{align}
    \Delta\epsilon_{\mathrm{sq}}    &=\Delta\epsilon\,=\,\tfrac{\nu}{2}\bigl(\cosh(2\rplus)-1\bigr).
\end{align}
Expressing $\rplus$ as a function of $\Delta\epsilon$ and $\nu$ then yields the result presented in Eq.~(\ref{eq:squeezing parameter rplus}), i.e., the worst precision for Gaussian single-mode unitaries is achieved when all energy is invested into single-mode squeezing.

\subsubsection{Minimal variance}\label{sec:min variance}

Next we are interested in determining the optimal strategy using Gaussian single-mode unitaries. To this end, we similarly define $\mathcal{V}_{-}:=4V_{-}/\omega^{2}$ with $V_{-}$ as in Eq.~(\ref{eq:VmaxVmin}), that is, we have to minimize
\begin{align}
    \mathcal{V}_{-} &=\nu^{2}\cosh(4r)\!-\!1\!+\!2\nu e^{-2r}\bigl( 2\Delta\epsilon\!-\!\nu\bigl[\cosh(2r)\!-\!1\bigr] \bigr)
    \nonumber\\[1mm]
    &=\,\nu^{2}\bigl[\sinh(4r)+2e^{-2r}(2\tfrac{\Delta\epsilon}{\nu}+1)\bigr]-(\nu^{2}+1).
    \label{eq:Vmin appendix}
\end{align}
The partial derivatives w.r.t. to $r$ yield
\begin{align}
    \frac{\partial\mathcal{V}_{-}}{\partial r}  &=\,
    4\nu^{2}\bigl[\cosh(4r)-(2\tfrac{\Delta\epsilon}{\nu}+1)e^{-2r}\bigr],
    \label{eq:partial deriv of V-}\\[1mm]
    \frac{\partial^{2}\mathcal{V}_{-}}{\partial r^{2}}  &=\,
    8\nu^{2}\bigl[2\sinh(4r)+(2\tfrac{\Delta\epsilon}{\nu}+1)e^{-2r}\bigr].
    \label{eq:second partial deriv of V-}
\end{align}
The extremal condition $\tfrac{\partial\mathcal{V}_{-}}{\partial r}|_{r=r_{\mathrm{extr.}}}=0$ then yields
\begin{align}
    (2\tfrac{\Delta\epsilon}{\nu}+1)  &=\,e^{2r_{\mathrm{extr.}}}\,\cosh(4r_{\mathrm{extr.}})\,,
    \label{eq:prec best case extr condition}
\end{align}
which has a unique solution $r_{\mathrm{extr.}}=\rminus$ for $r_{\mathrm{extr.}}\geq0$ since the function $e^{2r}\cosh(4r)$ is greater or equal than $1$ and is strictly increasing for $r\geq0$. Therefore, we have one and only one solution $\rminus\geq0$ for every value of $(2\tfrac{\Delta\epsilon}{\nu}+1)$. Moreover, inserting into the second derivative gives
\begin{align}
    \left.\frac{\partial^{2}\mathcal{V}_{-}}{\partial r^{2}}\right|_{r=\rminus}  &=\,
    4\nu^{2}\bigl[3e^{4\rminus}-e^{-4\rminus}\bigr],
\end{align}
which is positive for $\rminus>-\ln(3)/8$ and hence in particular when $\rminus>0$. Inserting $r_{\mathrm{extr.}}=\rminus$ into the condition of Eq.~(\ref{eq:prec best case extr condition}) and expressing $\Delta\epsilon$ one thus arrives at the result of Eq.~(\ref{eq:r implicit}), i.e.,
\begin{align}
    \Delta\epsilon  &=\,\tfrac{\nu}{2}\bigl(e^{2{\rminus} }\cosh(4{\rminus} )-1\bigr)\,.
\end{align}
Moreover, for the minimum we find that the energy input splits into squeezing and displacement according to
\begin{align}
    \Delta\epsilon_{\mathrm{sq}} &=\tfrac{\nu}{2} \bigl(\cosh(2\rminus)-1\bigr)\,,\\[1mm]
    \Delta\epsilon_{\mathrm{d}} &=\Delta\epsilon-\Delta\epsilon_{\mathrm{sq}}=
    \tfrac{\nu}{2} e^{4\rminus}\sinh(2\rminus)\geq0\,,
\end{align}
such that, unlike the maximum at $r_{\mathrm{extr.}}^{+}$ discussed before, the desired minimum can be physically realized for all $\Delta\epsilon$ and $\nu$.

\subsubsection{Maximal fluctuations}\label{sec:max fluctuations}

Let us now determine the maximal fluctuations that are possible during a charging process at fixed input energy via single-mode Gaussian unitaries. That is, we are interested in finding the maximum value of $(\Delta W_{+}(r)/\omega)^{2}$ from Eq.~(\ref{eq:DW max min}) over all $r$ for fixed $\Delta\epsilon$ and $\nu$. To simplify this task, we note that this is equivalent to the maximization problem for the function $\mathcal{W}_{+}(r)$, given by
\begin{align}
    \mathcal{W}_{+}(r)
    &=\tfrac{1}{4}\mathcal{V}_{+}-\tfrac{1}{2}(\nu-1)^{2}\cosh(2r),
    \label{eq:DW max min appendix}
\end{align}
which, up to terms independent of $r$, corresponds to $(\Delta W_{+}(r)/\omega)^{2}$ from Eq.~(\ref{eq:DW max min}). The first two partial derivatives w.r.t. $r$ are
\begin{align}
    &\frac{\partial\mathcal{W}_{+}}{\partial r}  \,=\,
    \frac{1}{4}\frac{\partial\mathcal{V}_{+}}{\partial r} - (\nu-1)^{2}\sinh(2r)\\[0.5mm]
    &\ =\nu^{2}\Bigl[ \bigl(2\tfrac{\Delta\epsilon}{\nu}+1\bigr)e^{2r}-\cosh(4r)-\bigl(\tfrac{\nu-1}{\nu}\bigr)^{2}\sinh(2r) \Bigr]\,,
    \nonumber\\[1mm]
    &\frac{\partial^{2}\mathcal{W}_{+}}{\partial r^{2}}  \,=\,
    \frac{1}{4}\frac{\partial^{2}\mathcal{V}_{+}}{\partial r^{2}} - 2(\nu-1)^{2}\cosh(2r)\\[0.5mm]
    &\ =2\nu^{2}\Bigl[ \bigl(2\tfrac{\Delta\epsilon}{\nu}+1\bigr)e^{2r}-2\sinh(4r)-\bigl(\tfrac{\nu-1}{\nu}\bigr)^{2}\cosh(2r) \Bigr]\,,
    \nonumber
\end{align}
where we have inserted for the partial derivatives of $\mathcal{V}_{+}$ from Eqs.~(\ref{eq:partial deriv of V+}) and~(\ref{eq:second partial deriv of V+}). For the purpose of solving the maximization problem, we introduce the notation $\chi:=\bigl(2\tfrac{\Delta\epsilon}{\nu}+1\bigr)\geq1$ and $\lambda:=(\nu-1)^{2}/\nu^{2}$, with $0\leq\lambda\leq1$ since $\nu\geq1$, such that the extremal condition $\bigl(\partial\mathcal{W}_{+}/\partial r\bigr)_{r=\rtplus}=0$ at the extremal point $r=\rtplus$ reads
\begin{align}
    \lambda\tfrac{1}{2}(1-e^{-4\rtplus})+\tfrac{1}{2}\bigl(e^{2\rtplus}+e^{-6\rtplus}\bigr) &=\,\chi.
    \label{eq:max fluc problem reformulated}
\end{align}
We then define $u:=e^{-2r}$ along with a family of functions $f_{\lambda}(u)$ via
\begin{align}
    f_{\lambda}(u)  &:=\lambda\tfrac{1}{2}\bigl(1-u^{2}\bigr)+\tfrac{1}{2}\bigl(\tfrac{1}{u}+u^{3}\bigr).
    \label{eq:flambda}
\end{align}
The maximization problem for $\mathcal{W}_{+}(r)$ can thus be formulated as the question: Does there exist a $u=u_{\chi}$, with $0<u_{\chi}\leq1$, for every pair of $\lambda$ and $\chi$, such that $f_{\lambda}(u=u_{\chi})=\chi$? To answer this question, we first determine the extremal point $u_{\lambda}$ of $f_{\lambda}$, i.e., such that
\begin{align}
    \left.\frac{\partial f_{\lambda}(u)}{\partial u}\right|_{u=u_{\lambda}} &=-\lambda u_{\lambda}-\tfrac{1}{2}\bigl(\tfrac{1}{u_{\lambda}^{2}}-3u_{\lambda}^{2}\bigr)\,=\,0,
    \label{eq:derivative of flambda}
\end{align}
which implies $g(u_{\lambda}):=\tfrac{1}{2}\bigl(3u_{\lambda}-\tfrac{1}{u_{\lambda}^{3}}\bigr)=\lambda$ for $u_{\lambda}>0$. Since $g(u_{\lambda})$ is a continuous, strictly increasing function of $u_{\lambda}$ that can take the values $g(u_{\lambda}=3^{-1/4})=0$ and $g(u_{\lambda}=1)=1$, there is exactly one $u_{\lambda}$ that satisfies $g(u_{\lambda})=\lambda$ for any $\lambda$ between $0$ and $1$, suggesting that $f_{\lambda}(u)$ always has a unique local extremal point within the interval $]0,1[$. Inspection of the second partial derivative, i.e.,
\begin{align}
    \left.\frac{\partial^{2} f_{\lambda}(u)}{\partial u^{2}}\right|_{u=u_{\lambda}} &=\Bigl(-\lambda-\tfrac{1}{u_{\lambda}^{3}}+3u_{\lambda}\Bigr)_{u=u_{\lambda}}\\
    &=\,\tfrac{1}{2}\bigl(3u_{\lambda}-\tfrac{1}{u_{\lambda}^{3}}\bigr)\,=\,\lambda\geq0,\nonumber
\end{align}
then reveals that the local extremum is a local minimum. Moreover, since $f_{\lambda}(u=1)=1$, this means that the minimal value is below one, $f_{\lambda}(u_{\lambda})<1$. In contrast, we have $f_{\lambda}(u)
\substack{\text{\tiny{$u\!\rightarrow\!0$}}\\ \longrightarrow}\infty$, suggesting that $f_{\lambda}(u)$ is strictly decreasing on the interval $[0,u_{\lambda}[$ and can there take any value between $f_{\lambda}(u_{\lambda})<1$ and $\infty$ (in particular, any value $\chi$). We have thus found that there is a unique $u_{\chi}<u_{\lambda}$ for every $\lambda$ and $\chi$ such that $f_{\lambda}(u_{\chi})=\chi$.

In other words, Eq.~(\ref{eq:max fluc problem reformulated}) has a unique solution $\rtplus$ for every valid $\lambda$ and $\chi$. To check that this solution is a maximum, we calculate
\begin{align}
    \hspace*{-4mm}&\left.\frac{\partial^{2}\mathcal{W}_{+}}{\partial r^{2}}\right|_{r=\rtplus} \!\!\!=\,
    2\nu^{2}\bigl[\chi e^{2\rtplus}-2\sinh(4\rtplus)-\lambda\cosh(2\rtplus)\bigr]\nonumber\\
    \hspace*{-4mm}&\ \ \ \ \ \ =\, 2\nu^{2}\bigl[\tfrac{\chi}{u_{\chi}}-\tfrac{1}{u_{\chi}^{2}}+u_{\chi}^{2}-\tfrac{\lambda}{2}\bigl(\tfrac{1}{u_{\chi}}+u_{\chi}\bigr)\bigr],
\end{align}
where we have substituted $e^{-2\rtplus}=u_{\chi}$. Further inserting for $\chi=f_{\lambda}(u_{\chi})$ from Eq.~(\ref{eq:flambda}), and comparing with Eq.~(\ref{eq:derivative of flambda}), we arrive at
\begin{align}
    \left.\frac{\partial^{2}\mathcal{W}_{+}}{\partial r^{2}}\right|_{r=\rtplus}  &=\,
    2\nu^{2}\Bigl[-\lambda u_{\chi}-\tfrac{1}{2}\bigl(\frac{1}{u_{\chi}^{2}}-3u_{\chi}^{2}\bigr)\Bigr]\nonumber\\
    &=\,2\nu^{2}\left.\frac{\partial f_{\lambda}(u)}{\partial u}\right|_{u=u_{\chi}}\,<\,0,
\end{align}
which is negative since $u_{\chi}<u_{\lambda}$ is below the minimum $u_{\lambda}$ of $f_{\lambda}(u)$. Consequently, the extremal value of $\mathcal{W}_{+}$ is a maximum, which we have thus shown exists and is unique for any fixed $\lambda$ and $\chi$ corresponding to fixed values of $\Delta E$ and $T$.

\subsubsection{Minimal fluctuations}\label{sec:min fluctuations}

Similarly, we now wish to determine the minimal fluctuations that are possible during a single-mode Gaussian unitary charging process at fixed input energy, i.e., we want to minimize $(\Delta W_{-}(r)/\omega)^{2}$ from Eq.~(\ref{eq:DW max min}) over all $r$ for fixed $\Delta\epsilon$ and $\nu$. As in the previous section, we simplify the problem by considering the equivalent minimization of the function $\mathcal{W}_{-}(r)$, given by
\begin{align}
    \mathcal{W}_{-}(r)
    &=\tfrac{1}{4}\mathcal{V}_{-}-\tfrac{1}{2}(\nu-1)^{2}\cosh(2r).
    \label{eq:DW max min appendix}
\end{align}
Up to terms independent of $r$, this function corresponds to $(\Delta W_{-}(r)/\omega)^{2}$ from Eq.~(\ref{eq:DW max min}). The first two partial derivatives w.r.t. $r$ are then
\begin{align}
    &\frac{\partial\mathcal{W}_{-}}{\partial r}  \,=\,
    \frac{1}{4}\frac{\partial\mathcal{V}_{-}}{\partial r} - (\nu-1)^{2}\sinh(2r)\\[0.5mm]
    &\ =\nu^{2}\Bigl[ -\bigl(2\tfrac{\Delta\epsilon}{\nu}+1\bigr)e^{-2r}+\cosh(4r)-\bigl(\tfrac{\nu-1}{\nu}\bigr)^{2}\sinh(2r) \Bigr]\,,
    \nonumber\\[1mm]
    &\frac{\partial^{2}\mathcal{W}_{-}}{\partial r^{2}}  \,=\,
    \frac{1}{4}\frac{\partial^{2}\mathcal{V}_{-}}{\partial r^{2}} - 2(\nu-1)^{2}\cosh(2r)\\[0.5mm]
    &\ =2\nu^{2}\Bigl[ \bigl(2\tfrac{\Delta\epsilon}{\nu}+1\bigr)e^{-2r}+2\sinh(4r)-\bigl(\tfrac{\nu-1}{\nu}\bigr)^{2}\cosh(2r) \Bigr]\,,
    \nonumber
\end{align}
where we have inserted for the partial derivatives of $\mathcal{V}_{-}$ from Eqs.~(\ref{eq:partial deriv of V-}) and~(\ref{eq:second partial deriv of V-}). We then proceed as in Appendix~\ref{sec:max fluctuations}, and formulate the extremal condition $\bigl(\partial\mathcal{W}_{-}/\partial r\bigr)_{r=\rtplus}=0$  at the extremal point $r=\rtminus$ in terms of the constants $\chi:=\bigl(2\tfrac{\Delta\epsilon}{\nu}+1\bigr)\geq1$ and $\lambda:=(\nu-1)^{2}/\nu^{2}$, with $0\leq\lambda\leq1$, as
\begin{align}
    \lambda\tfrac{1}{2}(1-e^{4\rtminus})+\tfrac{1}{2}\bigl(e^{-2\rtminus}+e^{6\rtminus}\bigr) &=\,\chi.
    \label{eq:min fluc problem reformulated}
\end{align}
To verify, that this condition can be met for all $\lambda$ and $\chi$, we again define a new variable $v:=e^{-2r}$ with $0\leq v\leq1$ for $r\geq0$, and define a family of functions $h_{\lambda}(v)$ via
\begin{align}
    h_{\lambda}(v)  &:=\lambda\tfrac{1}{2}\bigl(1-\tfrac{1}{v^{2}}\bigr)+\tfrac{1}{2}\bigl(v+\tfrac{1}{v^{3}}\bigr).
    \label{eq:hlambda}
\end{align}
The minimization problem for $\mathcal{W}_{-}(r)$ can thus be formulated as: Does there exist a $v=v_{\chi}$, with $0<v_{\chi}\leq1$, for every pair of $\lambda$ and $\chi$, such that $h_{\lambda}(v=v_{\chi})=\chi$? To provide an answer, we start again by determining if $h_{\lambda}$ has any extremal points in the allowed range of $v$. At such an extremal point $v=v_{\lambda}$ we would have
\begin{align}
    \left.\frac{\partial h_{\lambda}(v)}{\partial v}\right|_{v=v_{\lambda}} &=
    \lambda \tfrac{1}{v_{\lambda}^{3}}+\tfrac{1}{2}\bigl(1-3\tfrac{1}{v_{\lambda}^{4}}\bigr)\,=\,0,
    \label{eq:derivative of hlambda}
\end{align}
which would imply $\lambda=\tfrac{1}{2}\bigl(\tfrac{3}{v_{\lambda}}-v_{\lambda}^{3}\bigr)=:\tilde{g}(v_{\lambda})$. However, the function $\tilde{g}(v_{\lambda})$ is strictly decreasing for $v_{\lambda}\in[0,1]$, with the minimal value $\tilde{g}(v_{\lambda}=1)=1$. Therefore, $h_{\lambda}(v)$ has no local minima (or maxima) on the open interval $]0,1[$. Moreover, $h_{\lambda}(v)$ diverges as $v\rightarrow0$, and takes its minimum within the allowed range of $v$ for $v=1$, i.e., $h_{\lambda}(v=1)=1$. We have thus shown that $h_{\lambda}(v)$ is a strictly decreasing function of $v\in[0,1]$ that can take any value between $1$ and $\infty$. There is thus a unique value $v_{\chi}$ such that $h_{\lambda}(v=v_{\chi})=\chi$ for every $\chi$ and $\lambda$.

Finally, we check that we have indeed found a minimum of $\mathcal{W}_{-}$ by evaluating the second partial derivative at $r=\rtminus$ (corresponding to $v=v_{\chi}$), i.e.,
\begin{align}
    \hspace*{-4mm}&\left.\frac{\partial^{2}\mathcal{W}_{-}}{\partial r^{2}}\right|_{r=\rtminus} \!\!\!=\,
    2\nu^{2}\bigl[\chi e^{-2\rtminus}+2\sinh(4\rtminus)-\lambda\cosh(2\rtminus)\bigr]\nonumber\\
    \hspace*{-4mm}&\ \ \ \ \ \ =\, 2\nu^{2}\bigl[\chi v_{\chi}+\tfrac{1}{v_{\chi}^{2}}-v_{\chi}^{2}-\tfrac{\lambda}{2}\bigl(\tfrac{1}{v_{\chi}}+v_{\chi}\bigr)\bigr],
\end{align}
where we have substituted $e^{-2\rtminus}=v_{\chi}$. Inserting for $\chi=h_{\lambda}(v_{\chi})$ from Eq.~(\ref{eq:hlambda}), and comparing with Eq.~(\ref{eq:derivative of hlambda}), we obtain
\begin{align}
    \left.\frac{\partial^{2}\mathcal{W}_{-}}{\partial r^{2}}\right|_{r=\rtminus}  &=\,
    2\nu^{2}\Bigl[-\lambda \tfrac{1}{v_{\chi}}-\tfrac{1}{2}\bigl(v_{\chi}^{2}-3\frac{1}{v_{\chi}^{2}}\bigr)\Bigr]\nonumber\\
    &=\,2\nu^{2}v_{\chi}^{2}\left.\frac{\partial h_{\lambda}(v)}{\partial v}\right|_{v=v_{\chi}}\,\geq\,0,
\end{align}
which is nonnegative since $h_{\lambda}(v)$ is strictly decreasing for $0\leq v\leq1$, and hence has a negative first derivative on this interval. We can therefore finally conclude that the extremal value of $\mathcal{W}_{-}$ is a minimum that exists and is unique for any fixed $\lambda$ and $\chi$ corresponding to fixed values of $\Delta E$ and $T$.



\begin{thebibliography}{99}

\bibitem{GooldHuberRieraDelRioSkrzypczyk2016}
J.~Goold, M.~Huber, A.~Riera, L.~del~Rio, and P.~Skrzypczyk,\
\emph{The role of quantum information in thermodynamics \textemdash\ a topical review},\
\href{http://dx.doi.org/10.1088/1751-8113/49/14/143001}{J.\ Phys.\ A:\ Math.\ Theor.\ \textbf{49}, 143001 (2016)}\
[\href{http://arxiv.org/abs/1505.07835}{arXiv:1505.07835}].

\bibitem{MillenXuereb2016}
J.~Millen and A.~Xuereb,\
\emph{Perspective on quantum thermodynamics},\
\href{http://dx.doi.org/10.1088/1367-2630/18/1/011002}{New\ J.\ Phys.\ \textbf{18}, 011002 (2016)}\
[\href{http://arxiv.org/abs/1509.01086}{arXiv:1509.01086}].

\bibitem{VinjanampathyAnders2016}
S.~Vinjanampathy and J.~Anders,\
\emph{Quantum Thermodynamics},\
\href{http://dx.doi.org/10.1080/00107514.2016.1201896}{Contemp.\ Phys.\ \textbf{57}, 1 (2016)}\
[\href{http://arxiv.org/abs/1508.06099}{arXiv:1508.06099}].

\bibitem{BrandaoHorodeckiNgOppenheimWehner2015}
F.~G.~S.~L.~Brand{\~a}o, M.~Horodecki, N.~H.~Y.~Ng, J.~Oppenheim, and S.~Wehner,\
\emph{The second laws of quantum thermodynamics},\
\href{http://dx.doi.org/10.1073/pnas.1411728112}{Proc.\ Natl.\ Acad.}\
\href{http://dx.doi.org/10.1073/pnas.1411728112}{Sci.\ U.S.A.\ \textbf{112}, 3275 (2015)}\
[\href{http://arxiv.org/abs/1305.5278}{arXiv:1305.5278}].

\bibitem{BrandaoHorodeckiOppenheimRenesSpekkens2013}
F.~G.~S.~L.~Brand{\~a}o, M.~Horodecki, J.~Oppenheim, J.~M.~Renes, and R.~W.~Spekkens,\
\emph{The Resource Theory of Quantum States Out of Thermal Equilibrium},\
\href{http://dx.doi.org/10.1103/PhysRevLett.111.250404}{Phys.\ Rev.\ Lett.\ \textbf{111}, 250404}
\href{http://dx.doi.org/10.1103/PhysRevLett.111.250404}{(2013)}\
[\href{http://arxiv.org/abs/1111.3882}{arXiv:1111.3882}].

\bibitem{Mueller2017}
M.~P.~M{\"u}ller,\
\emph{Correlating thermal machines and the second law at the nanoscale},\
e-print \href{https://arxiv.org/abs/1707.03451}{arXiv:1707.03451} [quant-ph] (2017).

\bibitem{GogolinEisert2016}
C.~Gogolin and J.~Eisert,\
\emph{Equilibration, thermalisation, and the emergence of statistical mechanics in closed quantum systems},\
\href{http://dx.doi.org/10.1088/0034-4885/79/5/056001}{Rep.\ Prog.\ Phys.}\
\href{http://dx.doi.org/10.1088/0034-4885/79/5/056001}{\textbf{79}, 056001 (2016)}\
[\href{http://arxiv.org/abs/1503.07538}{arXiv:1503.07538}].

\bibitem{PuszWoronowicz1978}
W.~Pusz and S.~L.~Woronowicz,\
\emph{Passive states and KMS states for general quantum systems},\
\href{http://dx.doi.org/10.1007/BF01614224}{Commun.\ Math.}\
\href{http://dx.doi.org/10.1007/BF01614224}{Phys.\ \textbf{58}, 273 (1978)}.

\bibitem{PerarnauLlobetHovhannisyanHuberSkrzypczykTuraAcin2015}
M.~Perarnau-Llobet, K.~V.~Hovhannisyan, M.~Huber, P.~Skrzypczyk, J.~Tura, and A.~Ac{\'i}n,\
\emph{Most energetic passive states},\
\href{http://dx.doi.org/10.1103/PhysRevE.92.042147}{Phys.\ Rev.\ E\ \textbf{92},}
\href{http://dx.doi.org/10.1103/PhysRevE.92.042147}{042147 (2015)}\
[\href{http://arxiv.org/abs/1502.07311}{arXiv:1502.07311}].

\bibitem{BrownFriisHuber2016}
E.~G.~Brown, N.~Friis, and M.~Huber,\
\emph{Passivity and practical work extraction using Gaussian operations},\
\href{http://dx.doi.org/10.1088/1367-2630/18/11/113028}{New\ J.\ Phys.\ \textbf{18}, 113028 (2016)}\
[\href{http://arxiv.org/abs/1608.04977}{arXiv:1608.04977}].

\bibitem{PerryCwiklinskiAndersHorodeckiOppenheim2015}
C.~Perry, P.~{\'C}wikli{\'n}ski, J.~Anders, M.~Horodecki, and J.~Oppenheim,\
\emph{A sufficient set of experimentally implementable thermal operations},\
e-print \href{https://arxiv.org/abs/1511.06553}{arXiv:1511.06553} [quant-ph] (2017).

\bibitem{LostaglioAlhambraPerry2017}
{M.~Lostaglio, {\'A}.~M.~Alhambra, and C.~Perry},\
\emph{Elementary Thermal Operations},\
\href{http://dx.doi.org/10.22331/q-2018-02-08-52}{Quantum \textbf{2}, 52}
\href{http://dx.doi.org/10.22331/q-2018-02-08-52}{(2018)}\
[\href{https://arxiv.org/abs/1607.00394}{arXiv:1607.00394}].

\bibitem{MazurekHorodecki2017}
P.~Mazurek and M.~Horodecki,\
\emph{Decomposability and Convex Structure of Thermal Processes},\
e-print \href{https://arxiv.org/abs/1707.06869}{arXiv:1707.06869} [quant-ph] (2017).

\bibitem{ClivazSilvaHaackBohrBraskBrunnerHuber2017}
{F.~Clivaz, R.~Silva, G.~Haack, J.~Bohr Brask, N.~Brunner, and M.~Huber,}\
\emph{Unifying paradigms of quantum refrigeration: resource-dependent limits},\
e-print \href{https://arxiv.org/abs/1710.11624}{arXiv:1710.11624} [quant-ph] (2017).

\bibitem{HorodeckiOppenheim2013b}
{M.~Horodecki and J.~Oppenheim,}\
\emph{Fundamental limitations for quantum and nanoscale thermodynamics},\
\href{http://dx.doi.org/10.1038/ncomms3059}{Nat.\ Commun.\ \textbf{4}, 2059 (2013)}\
[\href{http://arxiv.org/abs/1111.3834}{arXiv:1111.3834}].

\bibitem{GourMuellerNarasimhacharSpekkensYungerHalpern2015}
{G.~Gour, M.~P.~M{\"u}ller, V.~Narasimhachar, R.~W.~Spekkens, and N.~Yunger~Halpern,}\
\emph{The resource theory of informational nonequilibrium in thermodynamics},\
\href{http://dx.doi.org/10.1016/j.physrep.2015.04.003}{Phys.\ Rep.\ \textbf{583}, 1-58 (2015)}\
[\href{https://arxiv.org/abs/1309.6586}{arXiv:1309.6586}].

\bibitem{Aberg2014}
{J.~{\AA}berg,}\
\emph{Catalytic Coherence},\
\href{http://dx.doi.org/10.1103/PhysRevLett.113.150402}{Phys.\ Rev.\ Lett.}\
\href{http://dx.doi.org/10.1103/PhysRevLett.113.150402}{\textbf{113}, 150402 (2014)},\
[\href{https://arxiv.org/abs/1304.1060}{arXiv:1304.1060}].

\bibitem{MalabarbaShortKammerlander2015}
{A.~S.~L.~Malabarba, A.~J.~Short, and P.~Kammerlander,}\
\emph{Clock-Driven Quantum Thermal Engines},\
\href{http://dx.doi.org/10.1088/1367-2630/17/4/045027}{New\ J.\ Phys.\ \textbf{17}, 045027 (2015)}\
[\href{https://arxiv.org/abs/1412.1338}{arXiv:1412.1338}].


\bibitem{SkrzypczykShortPopescu2013}
{P.~Skrzypczyk, A.~J.~Short, and S.~Popescu,}\
\emph{Extracting work from quantum systems},\
e-print \href{https://arxiv.org/abs/1302.2811}{arXiv:1302.2811} [quant-ph] (2013).

\bibitem{SkrzypczykShortPopescu2014}
{P.~Skrzypczyk, A.~J.~Short, and S.~Popescu,}\
\emph{Work extraction and thermodynamics for individual quantum systems},\
\href{http://dx.doi.org/10.1038/ncomms5185}{Nat.\ Commun.\ \textbf{5}, 4185}
\href{http://dx.doi.org/10.1038/ncomms5185}{(2014)}\
[\href{http://arxiv.org/abs/1307.1558}{arXiv:1307.1558}].

\bibitem{BinderVinjanampathyModiGoold2015}
F.~C.~Binder, S.~Vinjanampathy, K.~Modi, and J.~Goold,\
\emph{Quantacell: Powerful charging of quantum batteries},\
\href{http://dx.doi.org/10.1088/1367-2630/17/7/075015}{New\ J.\ Phys.\ \textbf{17}, 075015 (2015)}\
[\href{http://arxiv.org/abs/1503.07005}{arXiv:1503.07005}].

\bibitem{CampaioliPollockBinderCeleriGooldVinjanampathyModi2017}
F.~Campaioli, F.~A.~Pollock, F.~C.~Binder, L.~C.~C{\'e}leri, J.~Goold, S.~Vinjanampathy, and K.~Modi,\
\emph{Enhancing the charging power of quantum batteries},\
\href{https://doi.org/10.1103/PhysRevLett.118.150601}{Phys.\ Rev.\ Lett.\ \textbf{118}, 150601}
\href{https://doi.org/10.1103/PhysRevLett.118.150601}{(2017)}\
[\href{https://arxiv.org/abs/1612.04991}{arXiv:1612.04991}].

\bibitem{FerraroCampisiAndolinaPellegriniPolini2018}
{D.~Ferraro, M.~Campisi, G.~M.~Andolina, V.~Pellegrini, and M.~Polini,}\
\emph{High-Power Collective Charging of a Solid-State Quantum Battery},\
\href{http://dx.doi.org/10.1103/PhysRevLett.120.117702}{Phys.\ Rev.\ Lett.\ \textbf{120}, 117702 (2018)}\
[\href{https://arxiv.org/abs/1707.04930}{arXiv:1707.04930}].

\bibitem{HoferSouquetClerk2016}
{P.~P.~Hofer, J.-R.~Souquet, and A.~A.~Clerk,}\
\emph{Quantum heat engine based on photon-assisted Cooper pair tunneling},\
\href{http://dx.doi.org/10.1103/PhysRevB.93.041418}{Phys.\ Rev.\ B\ \textbf{93}, 041418}
\href{http://dx.doi.org/10.1103/PhysRevB.93.041418}{(2016)}\
[\href{https://arxiv.org/abs/1512.02165}{arXiv:1512.02165}].

\bibitem{HoferPerarnauLlobetBohrBraskSilvaHuberBrunner2016}
{P.~P.~Hofer, M.~Perarnau-Llobet, J.~Bohr Brask, R.~Silva, M.~Huber, and N.~Brunner,}\
\emph{Autonomous Quantum Refrigerator in a Circuit-QED Architecture Based on a Josephson Junction},\
\href{http://dx.doi.org/10.1103/PhysRevB.94.235420}{Phys.}\
\href{http://dx.doi.org/10.1103/PhysRevB.94.235420}{Rev.\ B\ \textbf{94}, 235420 (2016)}\
[\href{https://arxiv.org/abs/1607.05218}{arXiv:1607.05218}].

\bibitem{MitchisonHuberPriorWoodsPlenio2016}
{M.~T.~Mitchison, M.~Huber, J.~Prior, M.~P.~Woods, and M.~B.~Plenio,}\
\emph{Realising a quantum absorption refrigerator with an atom-cavity system},\
\href{http://dx.doi.org/10.1088/2058-9565/1/1/015001}{Quantum\ Sci.\ Technol.\ \textbf{1},}
\href{http://dx.doi.org/10.1088/2058-9565/1/1/015001}{015001 (2016)}\
[\href{https://arxiv.org/abs/1603.02082}{arXiv:1603.02082}].

\bibitem{MaslennikovDingHablutzelGanRouletNimmrichterDaiScaraniMatsukevich2017}
{G.~Maslennikov, S.~Ding, R.~Hablutzel, J.~Gan, A.~Roulet, S.~Nimmrichter, J.~Dai, V.~Scarani, and D.~Matsukevich,}\
\emph{Quantum absorption refrigerator with trapped ions},\
e-print \href{https://arxiv.org/abs/1702.08672}{arXiv:1702.08672} [quant-ph] (2017).

\bibitem{RossnagelDawkinsTolazziAbahLutzSchmidtKalerSinger2016}
{J.~Ro{\ss}nagel, S.~T.~Dawkins, K.~N.~Tolazzi, O.~Abah, E.~Lutz, F.~Schmidt-Kaler, and K.~Singer,}\
\emph{A single-atom heat engine},\
\href{http://dx.doi.org/10.1126/science.aad6320}{Science}\
\href{http://dx.doi.org/10.1126/science.aad6320}{\textbf{352}, 325 (2016)}\
[\href{https://arxiv.org/abs/1510.03681}{arXiv:1510.03681}].

\bibitem{Weedbrooketal2012}
C.~Weedbrook, S.~Pirandola, R.~Garc{\'i}a-Patr\'{o}n, N.~J.~Cerf, T.~C.~Ralph, J.~H.~Shapiro, and S.~Lloyd,\
\emph{Gaussian quantum information},\
\href{http://dx.doi.org/10.1103/RevModPhys.84.621}{Rev.}\
\href{http://dx.doi.org/10.1103/RevModPhys.84.621}{Mod.\ Phys.\ \textbf{84}, 621 (2012)}\
[\href{http://arxiv.org/abs/1110.3234}{arXiv:1110.3234}].

\bibitem{CampisiHaenggiTalkner2011}
{M.~Campisi, P.~H{\"a}nggi, and P.~Talkner,}\
\emph{Colloquium. Quantum Fluctuation Relations: Foundations and Applications},\
\href{http://dx.doi.org/10.1103/RevModPhys.83.771}{Rev.\ Mod.\ Phys.\ \textbf{83},}
\href{http://dx.doi.org/10.1103/RevModPhys.83.771}{771 (2011)};
Erratum: \href{https://doi.org/10.1103/RevModPhys.83.1653}{Rev.\ Mod.\ Phys.\ \textbf{83}, 1653}
\href{https://doi.org/10.1103/RevModPhys.83.1653}{(2011)}\
[\href{https://arxiv.org/abs/1012.2268}{arXiv:1012.2268}].

\bibitem{AlhambraMasanesOppenheimPerry2016}
{{\'A}.~M.~Alhambra, L.~Masanes, J.~Oppenheim, and C.~Perry,}\
\emph{The second law of quantum thermodynamics as an equality},\
\href{http://dx.doi.org/10.1103/PhysRevX.6.041017}{Phys.\ Rev.\ X\ \textbf{6}, 041017}
\href{http://dx.doi.org/10.1103/PhysRevX.6.041017}{(2016)}\
[\href{https://arxiv.org/abs/1601.05799}{arXiv:1601.05799}].

\bibitem{RichensMasanes2016}
{J.~G.~Richens and L.~Masanes,}\
\emph{From single-shot to general work extraction with bounded fluctuations in work},\
\href{http://dx.doi.org/10.1038/ncomms13511}{Nat.\ Commun.\ \textbf{7}, 13511 (2016)}\
[\href{https://arxiv.org/abs/1603.02417}{arXiv:1603.02417}].

\bibitem{EspositoHarbolaMukamel2009}
M.~Esposito, U.~Harbola, and S.~Mukamel,\
\emph{Nonequilibrium fluctuations, fluctuation theorems, and counting statistics in quantum systems},\
\href{https://dx.doi.org/10.1103/RevModPhys.81.1665}{Rev.\ Mod.\ Phys.\ \textbf{81}, 1665 (2009)}\
[\href{https://arxiv.org/abs/0811.3717}{arXiv:0811.3717}].

\bibitem{Olivares2012}
S.~Olivares,\
\emph{Quantum optics in the phase space - A tutorial on Gaussian states},\
\href{http://dx.doi.org/10.1140/epjst/e2012-01532-4}{Eur.\ Phys.\ J.\ \textbf{203},}
\href{http://dx.doi.org/10.1140/epjst/e2012-01532-4}{3 (2012)}\
[\href{http://arxiv.org/abs/1111.0786}{arXiv:1111.0786}].

\bibitem{Braunstein2005}
S.~L.~Braunstein,\
\emph{Squeezing as an irreducible resource},\
\href{http://dx.doi.org/10.1103/PhysRevA.71.055801}{Phys.\ Rev.\ A\ \textbf{71}, 055801 (2005)}\
[\href{http://arxiv.org/abs/quant-ph/9904002}{arXiv:quant-ph/9904002}].

\bibitem{LloydBraunstein1999}
S.~Lloyd and S.~L.~Braunstein,\
\emph{Quantum computation over continuous variables},\
\href{http://dx.doi.org/10.1103/PhysRevLett.82.1784}{Phys.\ Rev.\ Lett.}\
\href{http://dx.doi.org/10.1103/PhysRevLett.82.1784}{\textbf{82}, 1784 (1999)}\
[\href{http://arxiv.org/abs/quant-ph/9810082}{arXiv:quant-ph/9810082}].

\bibitem{BruschiPerarnauLlobetFriisHovhannisyanHuber2015}
D.~E.~Bruschi, M.~Perarnau-Llobet, N.~Friis, K.~V.~Hovhannisyan, and M.~Huber,\
\emph{The thermodynamics of creating correlations: Limitations and optimal protocols},\
\href{http://dx.doi.org/10.1103/PhysRevE.91.032118}{Phys.\ Rev.\ E\ \textbf{91}, 032118}
\href{http://dx.doi.org/10.1103/PhysRevE.91.032118}{(2015)}\
[\href{http://arxiv.org/abs/1409.4647}{arXiv:1409.4647}].

\bibitem{BruschiFriisFuentesWeinfurtner2013}
D.~E.~Bruschi, N.~Friis, I.~Fuentes, and S.~Weinfurtner,\
\emph{On the robustness of entanglement in analogue gravity systems},\
\href{http://dx.doi.org/10.1088/1367-2630/15/11/113016}{New\ J.\ Phys.\ \textbf{15},}
\href{http://dx.doi.org/10.1088/1367-2630/15/11/113016}{113016 (2013)}\
[\href{http://arxiv.org/abs/arXiv:1305.3867}{arXiv:1305.3867}].

\bibitem{PerarnauLlobetHovhannisyanHuberSkrzypczykBrunnerAcin2015}
M.~Perarnau-Llobet, K.~V.~Hovhannisyan, M.~Huber, P.~Skrzypczyk, N.~Brunner, and A.~Ac{\'i}n,\
\emph{Extractable work from correlations},\
\href{http://dx.doi.org/10.1103/PhysRevX.5.041011}{Phys.\ Rev.\ X\ \textbf{5}, 041011 (2015)}\
[\href{http://arxiv.org/abs/1407.7765}{arXiv:1407.7765}].

\bibitem{HuberPerarnauHovhannisyanSkrzypczykKloecklBrunnerAcin2015}
M.~Huber, M.~Perarnau-Llobet, K.~V.~Hovhannisyan, P.~Skrzypczyk, C.~Kl{\"o}ckl, N.~Brunner, and A.~Ac{\'i}n,\
\emph{Thermodynamic cost of creating correlations},\
\href{http://dx.doi.org/10.1088/1367-2630/17/6/065008}{New\ J.\ Phys.\ \textbf{17}, 065008 (2015)}\
[\href{http://arxiv.org/abs/1404.2169}{arXiv:1404.2169}].

\bibitem{FriisHuberPerarnauLlobet2016}
N.~Friis, M.~Huber, and M.~Perarnau-Llobet,\
\emph{Energetics of correlations in interacting systems},\
\href{http://dx.doi.org/10.1103/PhysRevE.93.042135}{Phys.\ Rev.\ E\ \textbf{93}, 042135 (2016)}\
[\href{http://arxiv.org/abs/1511.08654}{arXiv:1511.08654}].

\bibitem{BrunelliGenoniBarbieriPaternostro2017}
{M.~Brunelli, M.~G.~Genoni, M.~Barbieri, and M.~Paternostro},\
\emph{Detecting Gaussian entanglement via extractable work},\
\href{http://dx.doi.org/10.1103/PhysRevA.96.062311}{Phys.\ Rev.\ A\ \textbf{96}, 062311 (2017)}\
[\href{https://arxiv.org/abs/1702.05110}{arXiv:1702.05110}].

\end{thebibliography}
\end{document}